\documentclass{article}
\usepackage{setspace}
\usepackage{amsmath,amsbsy,amstext}

\usepackage{amsfonts}
\usepackage{blkarray}
\usepackage{multirow}
\usepackage{amssymb}
\usepackage{textcomp}
\usepackage[left=25mm,top=25mm,right=25mm,bottom=25mm]{geometry}
\usepackage{graphicx}
\usepackage{subfigure}
\usepackage{tikz}
\usepackage{epic}
\usepackage{lscape}
\usepackage{array}
\usepackage[authoryear]{natbib}
\usepackage{url}
\usepackage{float}
\usepackage{booktabs}
\usepackage{rotating}
\usepackage{epstopdf}
\begin{document}
\newcommand{\ra}[1]{\renewcommand{\arraystretch}{#1}}
\newcommand{\ip}[1]{{\left\langle #1 \right\rangle}}
\newcommand{\argmin}{\operatornamewithlimits{argmin}}\newtheorem{acknowledgment}{Acknowledgment}
\newtheorem{algorithm}{Algorithm}
\newtheorem{axiom}{Axiom}
\newtheorem{conclusion}{Conclusion}
\newtheorem{conjecture}{Conjecture}
\newtheorem{corollary}{Corollary}
\newtheorem{criterion}{Criterion}
\newtheorem{definition}{Definition}
\newtheorem{example}{Example}
\newtheorem{exercise}{Exercise}
\newtheorem{lemma}{Lemma}
\newtheorem{notation}{Notation}
\newtheorem{problem}{Problem}
\newtheorem{proposition}{Proposition}
\newtheorem{remark}{Remark}
\newtheorem{solution}{Solution}
\newtheorem{summary}{Summary}
\newtheorem{theorem}{Theorem}
\newcommand\independent{\protect\mathpalette{\protect\independenT}{\perp}}
\def\independenT#1#2{\mathrel{\rlap{$#1#2$}\mkern2mu{#1#2}}}
\bibliographystyle{Chicago}
\title{Stable and predictive functional domain selection with application to brain images}
\author{Ah Yeon Park\thanks{Statistical Laboratory, University of Cambridge, ayp21@cam.ac.uk},
John A. D. Aston\thanks{Statistical Laboratory, University of Cambridge, J.Aston@statslab.cam.ac.uk},
Fr\'ed\'eric Ferraty\thanks{Institut de Math\'ematiques de Toulouse, ferraty@math.univ-toulouse.fr}
}
\providecommand{\keywords}[1]{\textbf{\textit{Keywords---}} #1}
\maketitle
\begin{abstract}
Motivated by increasing trends of relating brain images to a clinical outcome of interest, we propose a functional domain selection (FuDoS) method that effectively selects subregions of the brain associated with the outcome. View each individual's brain as a 3D functional object, the statistical aim is to distinguish the region where a regression coefficient $\beta(t)=0$ from $\beta(t)\neq0$, where $t$ denotes spatial location. FuDoS is composed of two stages of estimation. We first segment the brain into several small parts based on the correlation structure. Then, potential subsets are built using the obtained segments and their predictive performance are evaluated to select the best subset, augmented by a stability selection criterion. We conduct extensive simulations both for 1D and 3D functional data, and evaluate its effectiveness in selecting the true subregion. We also investigate predictive ability of the selected stable regions. To find the brain regions related to cognitive ability, FuDoS is applied to the ADNI's PET data. Due to the induced sparseness, the results naturally provide more interpretable information about the relations between the regions and the outcome. Moreover, the selected regions from our analysis show high associations with the expected anatomical brain areas known to have memory-related functions.
\end{abstract}


\section{Introduction}
More than 25 million people in the world today suffer from dementia, mostly caused by Alzheimer's disease (AD). Indeed, the number of individuals affected is expected to significantly rise with a worldwide phenomenon of population ageing, as ageing is the greatest risk factor for the development of AD \citep{Evans89, Brookmeyer98,Bianchetti01,Brookmeyer11}. Specifically, an estimate shows that in 2050, approximately 80 million people will suffer from AD worldwide. In view of the current prevalence and the projection, the identification and validation of biomarkers for diagnosing AD and other forms of dementia are increasingly important. However, an accurate and early diagnosis of AD is difficult as early symptoms of the disease are shared by a variety of disorders, as reflected by their common neuropathological features \citep{Jacobs95,Nestor04,Swainson01,Humpel11}.

AD is a severe neurodegenerative disorder of the brain defined by loss of memory and cognitive decline. A probable diagnosis of AD can be established based on clinical criteria, including medical history, physical examination, laboratory tests, neuroimaging techniques and neuropsychological tests \citep{Khachaturian85,Nygaard03,Chong05,Vemuri08,Mckhann11}. In particular, many studies have shown that neuroimaging techniques can provide invaluable information about AD and are crucial for the early detection of AD \citep{Matsuda07, Ferreira11, Petrella03, Chetelat03, Mosconi07}. Moreover, preclinical AD is known to have an association with changes in both cognitive ability and brain images \citep{Caselli07, Caselli09, Twamley06}. Thus, finding the association between such measures can be of help for the early detection of AD. For example, \citet{Duchesne09} discovered a linear relationship between baseline magnetic resonance imaging (MRI) and a decline in cognitive ability after a year of scanning. Taking the association into consideration, when mild cognition impairment (MCI), or AD is diagnosed, one should combine neuroimaging techniques with neuropsychological tests that measure cognitive impairment to track progression of the illness and examine effectiveness of the treatment. To this end, in this article, we introduce a new statistical methodology, intended to select regions of the brain associated with cognitive decline. We further build a predictive model based on the selected region to predict cognitive ability of a new subject using his/her brain scan.

Structural MRI and metabolic positron emission tomography (PET) are the most clinically used and promising brain imaging techniques to detect abnormalities in individual brains which might be at risk for AD. Fludeoxyglucose (FDG) PET images analyzed in this article were acquired from the Alzheimer's Disease Neuroimaing Initiative (ADNI) database. A more detailed description of the initiative is explained in the supplementary material. Many statistical methods have been introduced to reveal a relation between brain images and a clinical outcome of interest. Univariate methods are intended to build a separate statistical model, either for each voxel, or for each region of interest (ROI). See for example, \citet{Herholz02}, \citet{worsley2002} and \citet{Lazar08}, among many others. In such univariate approaches, where a separate model is fitted for each voxel (or ROI), one must consider an appropriate adjustment to account for multiple comparisons and aggregate the results. Alternatively, in multivariate covariate approaches, every voxel is treated as a single predictor. Since the number of voxels is far larger than the number of images, ordinary least squares for standard linear regression cannot be employed without applying, for example, some regularization or dimension reduction.

Traditional univariate and multivariate approaches mentioned above share a common drawback: they do not consider important spatial information of the brain. To avoid loss of the information, new approaches incorporating the spatial structure have been introduced. For example, principal component analysis (PCA) can be carried out as dimension reduction, and then the selected principal components of the brain are used for further analysis \citep{Friston96, Kerrouche06, Caffo10}. Also, a wide range of Bayesian methods have been introduced \citep{Bowman08, Kang11, Zhang14}. In these Bayesian approaches, complex spatial correlation between voxels is modeled using appropriate prior distributions. More recently, the functional data analysis (FDA) framework has gained notable attention in the analysis of brain images. Functional data refers to the data whose visual representations are in functional forms, such as curves or images \citep{Ramsay05, Ferraty06}. \citet{Reiss10} used a functional version of PCA accounting for spatial features of the brain. Motivated by brain imaging studies on cognitive impairment in elderly subjects, \citet{Wang14} proposed 3D regularized functional regression, which accounts for the spatial information among neighboring voxels via Haar wavelets. \citet{Reiss15} introduced and compared several wavelet based predictive models to assess whether brain imaging data can predict presence or absence of attention deficit hyperactivity disorder (ADHD). Motivated by the complex structure of multidimensional brain image data, \citet{Zhou13} formulated a general tensor-based regression framework and applied the method to the ADHD data.

In this article, we take the functional view point and consider each individual's brain image as a 3D functional object, denoted by $X_i(t),i=1,...,n$, where $t$ refers to spatial location. Suppose $\mathbb{E}(Y)=\int X(t)\beta(t)dt$, where $Y$ is a clinical outcome of interest and $\beta(t)$ is a coefficient function or image. $\beta(t)$ can be thought of as a weight function, in that some parts of $\beta(t)$ with relatively large weights have a large influence on $Y$, while other parts with weights that are close to zero have nearly no impact. The brain is an complex object, consisting of a huge collection of small parts. Each of the parts has its own specific function, and very often they work together constructing a complicated relationship with each other. As functions of the brain are fairly regional, it would be natural to assume that only few parts of the brain are associated with a particular clinical outcome, and our goal is to find these specific subregions. However, most conventional estimation methods for $\beta(t)$ outlined in Section \ref{sec:lit review} do not allow $\beta(t)=0$ for some $t$, so identifying subregions where $\beta(t)\neq 0$ is not possible. We propose a new statistical methodology which, we call \textit{stable} and \textit{predictive} functional domain selection (FuDoS), which can effectively differentiate areas where $\beta(t)=0$ from $\beta(t)\neq 0$. Due to the sparse representation, the estimation result can naturally provide more interpretable information about the influence of such region of $\beta(t)\neq 0$. This information cannot be extracted from standard functional regression approaches, where the estimated $\beta(t)$ is continuous across the domain, and the effects at specific areas are ignored.

The article is outlined as follows. In Section \ref{sec:lit review}, we review some existing literature on the estimation of $\beta(t)$, including functional linear models and point of impact models. For ease of presentation, we explain and detail the proposed methodology with 1D functional data, then extend it to the 3D case in Section \ref{sec:extensions}. The two estimation stages, \textit{segmentation} and \textit{selection} are explained in Section \ref{sec:optimal segmentation} and Section \ref{sec:selection of segments}, respectively. A stable representation of the subdomain is determined in Section \ref{sec:stab selection}. Predictive models are built using the selected stable subregions in Section \ref{sec:constructing predictive model}. In Section \ref{sec:study 3d}, simulations for 3D functional data are given and finally the proposed method is applied to the analysis of ADNI's PET data. Section \ref{sec:conclusion} is the conclusion. In the supplementary material, we implement extensive simulations for 1D functional data, and illustrate the proposed method using the 1D gasoline data set.

\section{Stable and predictive functional domain selection (FuDoS)}
\label{sec:method}

\subsection{Review on the estimation of $\beta(t)$}
\label{sec:lit review}
We review some of the literature on functional regression, where the response $Y$ is scalar and the predictor $X$ is functional. Existing literature has mainly focused on linear models, where $Y$ is associated with $X$ through $\mathbb{E}(Y)=\int X(t)\beta(t)dt$. In practice, we only observe samples of a finite size, but the estimation target is intrinsically infinite dimensional which can give rise to identifiability issues. To deal with the problem, restrictions are often imposed on $\beta(t)$. Such restrictions generally involves a basis expansion with regularization, or penalization. In penalized B-splines approaches \citep{Marks99, Cardot03}, a penalty is often associated with a measure of the roughness, $\int\beta^{(d)}(t)^2dt$, where $\beta^{(d)}(t)$ denotes the $d$th order derivative, and $d=2$ is the most common choice. To account for sparsity and achieve variable selection, \citet{Zhao12} developed a wavelet-based lasso (WLasso) estimator, in which the subspace of $\beta(t)$ is restricted to the span of wavelets, and the wavelet coefficients are estimated via the lasso approach. It can result in sparse $\beta(t)$, but does not allow $\beta(t)$ to be exactly zero.

Most conventional estimation methods for $\beta(t)$, including those methods mentioned above, are not be able to identify regions of $\beta(t)=0$. Moreover, these methods often produce wiggly estimates that are hard to interpret. To aid interpretation, \citet{James09} developed a new methodology, called functional linear regression that's interpretable (FLiRTI). In this approach, first, one digitizes the domain into a fine set of points, and determine whether the $d$th derivatives of $\beta(t)$, $d=1,2..$, is zero or not at each point. The method is flexible in terms of the shape of $\beta(t)$. Also it can produce a highly interpretable estimate.

We also review some methods that concern the identification of points of impact. \citet{Ferraty10} proposed a method to detect predictive points of a predictor. The method is built around a stepwise forward algorithm, which selects a sequence of points giving the best prediction for $Y$. Although the algorithm works reasonably well, it does not explicitly account for the functional nature, treating each point as a separate predictor. Another possible limitation would be that it is unable to explore all possible combinations of points due to the nested nature of the algorithm. To deal with the second limitation, \citet{Ferraty14} advanced the forward selection and proposed a new methodology, called nonparametric variable selection (NOVAS). It enlarges the class of possible combinations of predictors while keeping the computational cost in a reasonable range. It is based on nonparametric regression, so it can take interaction effects between variables into account. To our knowledge, their algorithm has not been extended to the FDA context, and in this article we aim to investigate the extension.

A different approach for point of impact was introduced by \citet{McKeague10}. In their approach, $X$ is assumed to be a fractional Brownian motion with a Hurst parameter $H$, and the sensitive point, say $t^*$, is estimated based on a least squares approach. It is shown that the least-squares estimators of the points are consistent, and the convergence rate rises as $X$ gets more ragged, i.e., as $H$ gets smaller. \citet{Kneip15} generalized the point impact model, which incorporating both global and local effects. The key assumption ensuring the identifiability of those two effects is that the process exhibits specific local variation. This implies that at least some part of $X$ in a small neighbourhood of $t^*$ is essentially uncorrelated with the remainder of the trajectories outside of the interval, where $t^*$ denotes a true impact point. It is emphasized that the identifiability of the model does not impose any restriction on the degree-of-smoothness of $X(t)$. However, it is clear that detection of points of impact will benefit from a highly local variability that generally goes along with the non-smoothness.

\subsection{FuDoS model representation}
\label{sec:model rep}
Denote $Y$ a centred scalar response, and $X$ a centred functional predictor, which can be thought of as a function $X(t)$, $t\in[0,T]$, where $t$ here refers to time or location. For the 3D case that we study later, $t$ becomes a 3D coordinate $(h,v,z)$. Suppose that only few parts of $X$ have an association with $Y$, and our aim is to find the regions where $\beta(t)\neq0$. Let $X_{[l]}=\left\{X(t): t\in(s_{l-1},s_l]\right\}$ be the $l$th segment of $X$ defined by two boundary points $s_{l-1}$ and $s_l$, with $s_0=0$ and $s_L=T$, where $l=1,2,...,L$ refers to a segment index. Given a set of boundary points $\left\{s_l\right\}_{l=0}^{L}$, we seek the best subset of segments, denoted by $X_{\mathcal{J}^*}$, where $\mathcal{J}^*=[\kappa_1, \kappa_2,...,\kappa_K]$ is a collection of segment indices of the best subset. Denote $\beta_{[l]}$ a piece of $\beta(t)$, corresponding to $X_{[l]}$, that is, $\beta_{[l]}=\left\{\beta(t), t\in[s_{l-1},s_l)\right\}$. Given the segments, FuDoS model is formally written as
\begin{eqnarray}
\label{eq:local linear model}
Y&=&\sum_{k=1}^K \ip{X_{[\kappa_k]},\beta_{[\kappa_k]}}+\epsilon,
\end{eqnarray}
where $\ip{f_1,f_2}=\int f_1(t)f_2(t)dt$, and $\epsilon$ is an i.i.d observational error, with mean zero and variance of $\sigma^2$. We can generalize the above model to binary or count data using a link function, but in this article we only focus on the case where $Y$ is continuous. Typically, the size of the best subset, $K$, is far smaller than the total number of segments $L$, so the model can serve as a functional version of variable selection, so we call functional domain selection.

\subsection{Segmentation of $X$}
\label{sec:optimal segmentation}
In some cases, segments can be naturally given. In most cases, however, they are unknown, and we must estimate them. When correlation of $X$ is high, the adjacent points are likely to have similar values, consequently, they can result in similar predictive performance. Additionally, the presence of high correlation can deteriorate selection algorithms, by selecting an incorrect subset. This tendency becomes more severe when samples of $Y$ are corrupted by a sizeable error, or a regression coefficient is sufficiently smooth. This issue is also addressed in the context of points of impact. For instance, \citet{McKeague10} state that the convergence rate of the estimators of points of impact decreases when $X$ gets smoother. So the existence of high local variability can be helpful when trying to find points of impact \citep{Kneip15}. Taking empirical and the two theoretical results into consideration, we divide $X$ into several parts based on the correlation.

Our segmentation procedure is sequential. Denote $C(t,t')$ an absolute value of correlation between $X(t)$ and $X(t')$, that is, $C(t,t')=|\mathbb E[X(t)X(t')]|/\sqrt{G_X(t)}\sqrt{G_X(t')}$, where $G_X(t)$ is covariance of $X$ at point $t$, i.e., $G_X(t)=\mathbb E[X(t)X(t)]$. And define $I(s,u)=\iint_{[s,u]^2} \, C(t,t') dtdt'$. In Step 1 of the segmentation, we select the boundary point $s\in[0,T]$ that minimizes a loss function as
\begin{eqnarray}
\label{eq:seg step1}
\min\limits_{s}\left\{I(0,T)-\frac{1}{w_{0,s}}I(0,s)-\frac{1}{w_{s,T}}I(s,T)\right\}^2,
\end{eqnarray}
where $w_{u_1,u_2}$ is the size of the corresponding segment in \%, i.e., $w_{u_1,u_2}=100\times|u_2-u_1|/T$. Here, the inverse $1/w_{.,.}$ serves as a normalizing constant. Denote the solution to (\ref{eq:seg step1}) by $s_{1}^*$, where the lower index indicates the step number of the procedure, and denote the corresponding minimized error by $U(s_1^*)=U_1^*$. Given $s_{1}^*$, in Step 2, we find the location that optimizes either
\begin{eqnarray}
\min\limits_{s<s_{1}^*}\left\{I(0,T)-\frac{1}{w_{0,s}}I(0,s)-\frac{1}{w_{s,s_{1}^*}}I(s,s_{1}^*)-\frac{1}{w_{s_{1}^*,T}}I(s_{1}^*,T)\right\},
\end{eqnarray}
or
\begin{eqnarray}
\min\limits_{s_{1}^*\leq s}\left\{I(0,T)-\frac{1}{w_{0,s_{1}^*}}I(0,s_{1}^*)-\frac{1}{w_{s_{1}^*,s}}I(s_{1}^*,s)-\frac{1}{w_{s,T}}I(s,T)\right\}.
\end{eqnarray}
Similarly, we write its solution as $s_{2}^*$ and the corresponding minimized error by $U_2^*$. A loss function of the subsequent steps is built in similar fashion: a new integral term is progressively added to the preceding loss function. We report some simulated segmentation results in the supplementary material. The algorithm seems to perform reasonably well as it splits region with low correlation, while keeping region with high correlation intact.

The segmentation procedure is essentially equivalent to approximating the complete correlation by its subdiagonal parts. Without a normalizing constant, i.e., setting $w_{.,.}=1$, the above loss function amounts to \textit{squared sum of off-subdiagonal parts} of $C$. In such case, the minimized approximation error $U_j^*$ where $j$ is the step number, would always get larger as the algorithm progresses. On the other hand, the normalizing constant can have an opposite effect as it penalizes a segment with large size. Typically, the minimized approximation error path $U_j^*$ is convex. When the error path begins to rise, we terminate the segmentation procedure. However, when $X$ is extremely rough, the path can constantly decrease, so the algorithm will produce a too complex segmentation. To regulate complexity of the segmentation, we introduce a penalty, as an increasing function of number of segments, to the loss function. This produces the following penalized loss function
\begin{equation}
\tilde{U}_j^*=U_j^*+\rho L_j,
\label{equ:seg_penalization}
\end{equation}
where $\rho$ is a tuning parameter that controls complexity of the segmentation, and $\tilde{U}_j^*$ is the approximation error penalized by the number of segments $L_j$ at $j$th step. We employ a subsampling scheme for selecting the amount of segmentation, as detailed in Section \ref{sec:stab selection}.

The underlying assumption imposed on the proposed segmentation procedure is that the correlation has a bandable-shape, i.e., entries of the correlation decay as they move away from diagonal. Such a correlation structure naturally arises in a wide range of cases, including temporal or spatial data, the two most common forms of functional data. When the shape of the correlation largely deviates from the standard bandable-shape, e.g., functional data with a periodic pattern, the above procedure may not produce a sensible segmentation as non-adjacent segments actually have higher correlation. In this case, one may alter the above to incorporate these types of structure, allowing the combination of correlated segments that are far apart.

\subsection{Identifying predictive subdomain of $X$}
\label{sec:selection of segments}
Having obtained the segments, we now seek the most predictive subset of segments. 
Denote $\mathcal{J}$ a subset of segment indices $r=1,...,L$, so that $X_{\mathcal{J}}$ means the subdomain of $X$ associated with segments included in $\mathcal{J}=\left\{l_1,...,l_J\right\}$. In other words, it is a collection of segments, i.e., $X_{\mathcal{J}}=\cup_{j=1}^J X_{[l_j]}$, with $X_{[l_j]}$ as defined earlier in Section \ref{sec:model rep}. In each step of the selection procedure, we build a sequence of distinct potential subsets, denoted by $\mathcal{J}_1,...,\mathcal{J}_R$, for different values of $R$, and perform prediction of $Y$ based on each subset. Once we obtain a measure of the predictive performance for each subset using a 5-fold cross-validated (CV) error, the subsets are ordered in an ascending order of the CV error. We denote the ranked subsets by $\mathcal{J}(1),\mathcal{J}(2)...,\mathcal{J}(R)$, and use them to construct a new sequence of subsets for the subsequent step. For instance, if the 5th segment has the smallest CV error, followed by $10, 4, 12$..., then $\mathcal{J}^1(1)=\{5\}$, $\mathcal{J}^1(2)=\{10\}$, $\mathcal{J}^1(3)=\{4\}$ and $\mathcal{J}^1(4)=\{12\}$..., where the upper index indicates the algorithm step number, and the number in round brackets means the rank. In Step 2, we merge the ordered sequence in a pairwise manner to create a new sequence of potential subsets, leading to $\mathcal{J}_1=\{5,10\}$, $\mathcal{J}_2=\{4,5\}$, $\mathcal{J}_3=\{5,12\}$,...$\mathcal{J}_{r_1}=\{4,10\}$.... Similar to Step 1, after performing the prediction based on each subset, $\mathcal{J}_1,...,\mathcal{J}_R$, the sets are ranked according to their predictive ability, producing $\mathcal{J}^2(1),...,\mathcal{J}^2(R)$. If $\mathcal{J}^2(1)=\{2,5\}$, $\mathcal{J}^2(2)=\{2,7\}$, $\mathcal{J}^2(3)=\{1,9\}$..., a new sequence of subsets for Step 3 are $\mathcal{J}_1=\{2,5,7\}$ and $\mathcal{J}_2=\{1,2,5,9\}$...., and so on. In Step 3, our procedure produces subsets of size 3 or 4. Again, we perform the prediction based on each subset, $\mathcal{J}_1,...,\mathcal{J}_R$, and rank the sets based on their predictive performance. We continue the selection until we attain the best subset, and terminate the algorithm when the subsequent minimal CV error does not satisfy a rule, as given in (\ref{c:thred}).

Naive use of the above sequential algorithm can raise a computational concern. Specifically, in Step 1, we search the most predictive single segment over $R=L$ sets, where $L$ is the total number of segments. In Step 2, we select the best combination of two segments, so we have to explore $R$ number of subsets, where $R=\binom L2=\frac{L!}{2!(L-2)!}$. From a computational view point, it is undesirable because even a moderate value of $L$ can produce a large $R$, e.g., $L=20$ leads to $R=190$. Inspired by \citet{Ferraty14}, we reduce computational labour by keeping only the first top $\sqrt q$ subsets when building the sets for the subsequent step, where $q$ is an unknown value depending on the capability of the computational resources. If a set of segments does not seem to be useful on its own, ranked as top $\sqrt q+1, \sqrt q+2,...$, it is less likely to be useful when working with other sets. In practice, the functional variable $X$ is never continuously observed over the whole interval $[0,T]$, but at a grid of measurements of size $p$. Unlike the NOVAS method \citep{Ferraty14}, we somehow already reduce the computational burden by focusing our attention on the previous $L$ subdomains of $X$ coming from our segmentation step (instead of the $p$ digitized points). When we used $q=L$ for 1D functional data, the result was not satisfactory since $L$ is already quite small. In our 1D numerical study presented in the supplementary material, $L$ is not larger than 20, and $p$ is only a few hundred. Based on our empirical experience, we suggest $q=p/2$ for 1D functional data, as rule of thumb. On the other hand, in the ADNI's PET brain image analysis that we study in Section \ref{sec:study 3d}, the involved number of voxels is approximately 2.5 million, so we use $q=L/2$. We note that in our brain image analysis, $\hat{L}$ ranges between 1290 and 5380. Using the suggested cut-off rule as above, we reduce the computational time without affecting the performance of the proposed method.

As we mentioned earlier, in each step of the selection algorithm, we rank the subsets based on their predictive ability, and for its quantification, we use a 5-fold CV. Specifically, we randomly divide $n$ pairs of samples $(X_i,Y_i)$ drawn from $(X,Y)$ into 5 roughly equal parts. Holding out each sample fold for use as a validating set, we train the model with the remaining samples, yielding $\hat{g}^{-j}(X_{\mathcal{J}_r})$, where $\hat{g}^{-j}(X_{\mathcal{J}_r})$ is the leave-one-out estimator exclusive of the $j$th sample fold based on the subdomain $X_{\mathcal{J}_r}$, and we use a linear form for $g$ as explained in the subsequent paragraph. Using $\hat{g}^{-j}(X_{\mathcal{J}_r})$, we perform prediction of $Y_i$, $i\in I_{j}$, where $I_{j}$ is the $j$th part of sample indices. Repeating this procedure for each $j=1,...,5$, we compute the mean squared prediction error of the $r$th subset as
\begin{equation}
\text{CV}_r= 1/5\sum_{j=1}^5\sum_{i\in I_{j}}^{n_j} 1/n_{j} \left\{Y_i-\hat{g}^{-j}(X_{\mathcal{J}_r})\right\}^2,\quad \sum_{j=1}^5n_{j}=n,
\label{eq:nfold cv}
\end{equation}
where $n_j$ is the number of samples in $I_{j}$. Once we obtain CV error for each potential subset, the sets are ordered in an ascending order of CV error as CV$(\mathcal{J}^k(1))\leq,...,\leq$ CV$(\mathcal{J}^k(R))$, where $k$ is the step number. As explained, the ordered subsets are merged in a pairwise manner and a new sequence of subsets is built for the next step. The selection algorithm is terminated at $K$th step when the successive minimum CV error satisfies the following criterion
\begin{equation}
\frac{\text{CV}^*_{K}-\text{CV}^*_{K+1}}{\text{CV}^*_{K}}\leq c,
\label{c:thred}
\end{equation}
where $\text{CV}^*_{K}$ is the minimum value of CV error at the $K$th selection step, that is, $\text{CV}^*_{K}=\text{CV}(\mathcal{J}^{K}(1))$, where $\mathcal{J}^{K}(1)$ denotes the most predictive subset at $K$th selection step. The unknown tuning parameter $c$ controls the degree of selection. When $c$ is large, the algorithm will stop early, leaving out potentially relevant segments. If $c$ is small, by contrast, the selected subset can include false segments as a result of over-fitting. We provide detailed discussion on the selection of $c$ in Section \ref{sec:stab selection}.

To fit the regression function $g$ in (\ref{eq:nfold cv}), we use a linear model as given in (\ref{eq:local linear model}). The form of $\beta_{[l]}$ is unknown, and is dependent upon applications. To gain flexibility, one may use a spline basis expansion with penalization \citep{Marks99, Cardot03}. Such penalized basis approach can provide more flexible control over the shape of $\beta$, with the shape being determined by data. For instance, when the relation between $X$ and $Y$ is linear over the subset $X_{\mathcal[l]}$, $\hat{\beta}_{[l]}$ will reflect the relation by choosing a large value for the smoothing parameter as found in our gasoline example, see Figure \ref{fig:gasoline fig} in the supplementary material. Using a complex form for fitting $\beta_{[l]}$ would not cause a serious problem, when the sample size is large enough. However, when it is small, compared to the dimension of the data, and brain image data is a typical example of such data set, a complex form is likely to result in over-fitting, and hence unstable estimation and poor prediction. Indeed, results in our 1D simulation study in Table \ref{tab:1D Tab1} and Table \ref{tab:1D Tab2} the supplementary material reveal that when $n<<p$, using penalized splines methods for fitting for each $\beta_{[l]}$ can lead to over-fitting as indicated by large values of prediction error.

\subsection{Selecting the stable subdomain via subsampling}
\label{sec:stab selection}
The proposed method involves two tuning parameters: 1) $\rho$ regulates the complexity of the segmentation; and 2) $c$ determines the point of termination of the selection procedure. They are interrelated and have a joint effect on the amount of selection. When selecting the best subset, the major concern is to determine whether there exists a pair $(\rho,c)$ that identifies the true subset with high probability, and the aim is to choose such a pair. Data-driven methods such as cross-validated approaches may provide the simplest tool for the selection. However, the best model chosen by cross-validation in the lasso, for instance, tends to include too many variables \citep{Meinshausen06, Leng06}, and in our analysis of brain data we observed that the selection can be specific to a dataset. We attempt to avoid the situation by combining our selection algorithm with a generic subsampling scheme. Specifically, instead of choosing a single set of tuning parameters to determine the best subset, we perturb the data many times and select regions that appear in selected subsets with high probability. The spirit is that the stable subdomain should be consistently identified on similar sets of data.

Motivated by \citet{Meinshausen10}, we define the selection probability of each subdomain and the stable subdomain as follows. The \textit{selection probability} of any subdomain $\mathcal{X}\subseteq X$ is the probability of being in $X^{{\rho,c}}_{\mathcal{J}^*}$, where $X^{\rho,c}_{\mathcal{J}^*}$ is the subdomain of $X$ associated with the selected segment index set $\mathcal{J}^*$, given $\rho$ and $c$. Recall that we introduced two stages of estimation for $X^{\rho,c}_{\mathcal{J}^*}$ in previous subsections. Let $S$ be a random subsample of $\left\{1,...,n\right\}$ of size $\lceil n/2 \rceil$, drawn without replacement, and we use the subsample to obtain $X^{{\rho,c}}_{\mathcal{J}^*}$. Here, the selected set is implicitly a function of $S$, so we incorporate this dependence by writing $X^{{\rho,c}}_{\mathcal{J}^*}(S)$. As introduced in Section \ref{sec:selection of segments}, our selection algorithm is based on minimizing a 5-fold cross validation, which displays an additional source of randomness to the selected set. So, we write $X_{\mathcal{J}^*}^{{\rho,c}}(S,I)$, where $I$ is a 5-fold random split of the subsample $S$. Mathematically, the selection probability of the subdomain $\mathcal{X}$ given $\rho$ and $c$ is defined as
\begin{equation}
\Phi^{\rho,c}_{\mathcal{X}}=P \left\{\mathcal{X} \subseteq X^{\rho,c}_{\mathcal{J}^*}(S,I)\right\},
\label{equ:selection proba}
\end{equation}
where the probability $P$ is in terms of two sources of randomness $S$ and $I$. The estimate of the above probability can be naturally estimated by repeating the subsampling procedure a large number of times, and computing the relative frequency for $\mathcal{X} \subseteq X^{{\rho,c}}_{\mathcal{J}^*}(S,I)$. Based on the estimated selection probability $\hat{\Phi}^{{\rho,c}}_{\mathcal{X}}$, we define \textit{stable subdomain} as
\begin{equation}
\mathcal{X}_{stable}^\pi=\left\{\mathcal{X}: \max_{({\rho,c})\in B} \hat{\Phi}^{{\rho,c}}_{\mathcal{X}}>\pi\right\},
\label{equ:cutoff}
\end{equation}
where $\pi$ is an user-defined cut-off probability and where $\rho$ (resp. $c$) belongs to some given grid of values.

Having used the stability selection procedure, the problem has shifted from the choice of $\rho$ and $c$ to the choice of $\pi$ and a two-dimensional grid $B$ for both tuning parameters $\rho$ and $c$. Choosing the optimal pair of $(\rho,c)$ is an extremely difficult problem in high dimensional settings, while subsampling can provide a more straightforward and general framework for the problem as choosing fewer subregions or increasing $\pi$ will reduce the expected rate of falsely selected subregion. A major advantage of stability selection would be that the choice of $B$ does not have a large effect on the result, as long as it is varied within reasonable limits \citep{Meinshausen10}. We find a similar effect as discussed below. Given $\rho$, decreasing $c$ \textit{tends to} select the subset in an incremental manner, i.e., $X^{\rho,c_1}_{\mathcal{J}^*}\subseteq X^{\rho,c_2}_{\mathcal{J}^*}$, for $c_{2}\leq c_{1}$. Of course, as our selection procedure is not nested in nature, the above incremental relation would not hold theoretically, but we found that when $c$ is fairly small, say $c\leq 0.05$, the above relation tends to be satisfied. Considering this issue, so-called pointwise control \citep{Meinshausen10}, we only consider a single value for $c$, i.e., $c=0.01$, in such a way that some over-fitting occurs, so each selected subset $X^{{\rho,c}}_{\mathcal{J}^*}$ would contain the true subset with high probability. Unlike $c$, as $\rho$ does not explicitly exhibit the above incremental relation, we consider the minimum and maximum number of segments, and vary values for $\rho$, in a way that $\hat{L}$ (the estimated number of segments) changes within this range.

We now give guideline on $\pi$. Unlike $B$, the choice of $\pi$ is more directly related to the selection results. As decreasing $\pi$ generally increases the size of $\hat{\mathcal{X}}_{stable}^\pi$, it is likely to include the true subset with high probability. Choosing a small $\pi$ however would increase the expected rate of falsely selected subregions. One possible way of choosing $\pi$ would be evaluating predictive performance of each $\hat{\mathcal{X}}_{stable}^\pi$, and selecting $\pi$ that yields the smallest prediction error. Another possibility would be monitoring the maximized selection probability for each subdomain $\mathcal{X}$, and searching if there is any clear threshold for the choice. For instance, in the analysis of gasoline data in the supplementary material, we find that two subregions clearly stand out with selection probability higher than 0.8, see a red line in Figure S\ref{fig:gasolineFig1}. A sensible choice between the two possible approaches would depend on the aim of analysis. We find that the best $\pi$, in the sense that it attains the best predictive result, tends to get smaller as the sample size $n$ gets smaller, or the size of observational error on $Y$ becomes larger. These two quantities interplay, but $n$ seems to have a stronger effect. We shall give more detailed discussion on the selection of $\pi$ in Section \ref{sec:study 3d} and in the supplementary material.

Stability selection has a very attractive theoretical property that, under some assumptions and model settings, a certain bound on the expected number of false selections is guaranteed \citep{Meinshausen10}. In this article, we do not investigate its theoretical properties, rather we only highlight the two practical advantages offered by the scheme. First, it increases the selection probability at boundaries of the true segments. Due to the nature of the proposed method, where segments are predetermined without considering its relation to $Y$, the selection probability at the true boundaries can be low, if the estimated segments do not coincide with the true segments. Using stability selection can add flexibility to the segmentation procedure, so that the estimated boundaries can move around at each repetition of subsampling. Our empirical results show that this procedure indeed overcomes the boundary issue. The second benefit is that it offers a nice tool for stabilizing the selected subset, reducing the rate of falsely selected subregion without compromising the predictive power. Our simulation study presented in the supplementary material reveals that FuDoS can yield comparatively good prediction performance for a range of $\pi$. Additionally, the rate of falsely selected subregion seems to be reasonably low. We shall emphasize the second advantage in Section \ref{sec:study 3d} as well as in the supplementary material.

\subsection{Building predictive models based on stable subdomain}
\label{sec:constructing predictive model}
One of the most important and popular goals in brain image studies is prediction of a disease, or a clinical outcome using brain image data. For this goal, we attempt to develop predictive models based on selected stable subregions as explained below. Assume that we have a sequence of the selected stable subdomains for different values of $\pi$, and $\mathcal{X}_{stable}^{\pi_1} \subseteq \mathcal{X}_{stable}^{\pi_2}$, for $\pi_2\leq \pi_1$. And let $\mathcal{T}_{stable}^{\pi}$ be the domain on which $\mathcal{X}_{stable}^{\pi}$ is defined. Then, for each $\pi$, a predictive model is built
\begin{equation}
M_{\pi}:\quad Y=g(\mathcal{X}_{stable}^{\pi}) + \epsilon.
\label{eq:predictive model}
\end{equation}
Throughout the paper, we consider linear models for fitting $g$, so (\ref{eq:predictive model}) becomes
\begin{eqnarray}
\label{eq:predictive linear model}
M_{\pi}:\quad Y&=&\ip{X, \beta}_{\mathcal{T}_{stable}^{\pi}} + \epsilon,\\ \nonumber
&=&\ip{X, \beta_{\mathcal{X}_{stable}^{\pi}}} + \epsilon, \quad
\end{eqnarray}
where $\beta_{\mathcal{X}_{stable}^{\pi}}$ is a regression coefficient with flat region, i.e., $\beta_{\mathcal{X}_{stable}^{\pi}}=0$, for $t\notin\mathcal{T}_{stable}^{\pi}$, and $\beta_{\mathcal{X}_{stable}^{\pi}}\neq 0$, for $t\in\mathcal{T}_{stable}^{\pi}$. Under this setting, the fitted curve, or image of $\beta$ is zero over the region where $t\notin\mathcal{T}_{stable}^{\pi}$.

To fit $\beta_{\mathcal{X}_{stable}^{\pi}}$, the same model as used to find $\mathcal{X}_{stable}^{\pi}$ is considered. For instance, when a penalized splines fitting criterion is used as in our 1D numerical study, a equi-spaced sequence of knots is placed over the subregion $\mathcal{T}_{stable}^{\pi}$, and the roughness of $\beta(t)$ is controlled by a smoothing parameter. While when piecewise constant basis is used as in our 3D numerical study, one must determine the size of each piece, which amounts to dividing $\mathcal{T}_{stable}^{\pi}$ into several parts, over each part a constant function is fitted. We divide $\mathcal{T}_{stable}^{\pi}$ into several pieces using a density based clustering algorithm for spatial data \citep{Ester96, Sander98}. The algorithm groups together points that are closely located, and marks points as outliers, when they locate alone in low density regions. However, as our aim here is not to identify outlying points among the selected points, but split them into several groups, merely based on their locations, we consider the outlying points forming groups with low density. Under our selection framework, although the density is low, the outlying points (whose nearest neighbors are far apart) in the selected stable subset would have predictive power with high probability, and their mean effects on $Y$ would be quite different from a big cluster of points, if they are far apart. The density based clustering algorithm provides appropriate tools for our problem. First, it does not require one to specify the number of clusters a priori, as opposed to K-means clustering \citep{Hartigan79}. Moreover, the algorithm works well when the shape of clusters is arbitrary. It is efficient to implement and almost deterministic. For the implementation, we used the R-function $\textit{dbscan}$ in R-package $\textit{dbscan}$ \citep{Rpackage.dbscan}. As explained, when prediction of $Y$ is the purpose of analysis, $\pi$ can be selected by evaluating the predictive performance of each $M_{\pi}$, and selecting $\pi$ that yields the smallest prediction error. We find that the predictive performance varies little for a range of $\pi$. We give more detailed discussion on the selection in Section  \ref{sec:study 3d} and in the supplementary material.

\subsection{Extensions to 3D}
\label{sec:extensions}
We treat each individual's brain image as a 3D functional object and therefore extend the proposed methodology to 3D functional data. Essentially, for the selection, the same algorithm is used, regardless of the dimensionality. However, segmenting a multi-dimensional functional object is more complicated and computationally challenging. Unlike the 1D case, where segments are given as non-overlapping intervals (and are in some sense totally ordered), segments of the brain are non-overlapping 3D volumes in various shapes. To simplify the problem we assume that the complete 6-dimensional covariance function of the brain admits a separable form. The separability largely reduces the set space, over which we explore to find the optimal segmentation, leading to a significant increase in computational speed. For example, without the separability, the dimension of the set space that has to be explored in the ADNI's PET data is approximately 2.5 million $(\approx160\times 160\times 96)$. While when the separability is assumed, the size of the set space shrinks approximately to 410 $(\approx160+160+96)$. Of course, we gain computational efficiency at a price of precision. The use of separable functions for brain image data is not new, we refer the reader for instance to \citet{Aston12}.

We now detail the separability. Denote $X_i(h,v,z)$ the $i$th sample of $X(h,v,z)$, where $h\in H$, $v\in V$ and $z\in Z$ represent voxel location in the brain, with $H$, $V$ and $Z$ being compact sets. We can translate the 3D brain into a 1D functional object as $X(t)\equiv X(h,v,z)$, where $t\equiv (h,v,z)$. We assume that $X$ is centred and denote the spatial full covariance function of $X(h,v,z)$ by $G\Big(\big(h,v,z\big),\big(h',v',z'\big)\Big)=\mathbb E[X\big(h,v,z\big)X\big(h',v',z'\big)]$. Suppose that $G$ has the following separable form
\begin{equation}
G\Big(\big(h,v,z\big),\big(h',v',z'\big)\Big)=G_H(h,h')G_V(v,v')G_Z(z,z'),
\label{eq:separability}
\end{equation}
where $G_H(h,h')$, $G_V(v,v')$ and $G_Z(z,z')$ are marginal covariance with $G_H$ (resp. $G_V$ and $G_Z$) mapping $H\times H$ (resp. $V \times V$ and $Z \times Z$) into $\mathbb{R}$. To obtain $G_H$, $G_V$ and $G_Z$ we follow the calculation, as used in \citet{Aston12}. To obtain $G_H$, we integrate out $G$ with respect to $V$ and $Z$ as
\begin{equation}
G_H(h,h')=\int_S\int_VG\Big(\big(h,v,z\big),\big(h',v,z\big)\Big)dvdz,
\label{eq:marginal cov}
\end{equation}
and its estimate $\hat{G}_H(h,h')$ is computed by replacing $G\Big(\big(h,v,z\big),\big(h',v',z'\big)\Big)$ with its sample alternative as
\begin{equation}
\resizebox{.7 \textwidth}{!}
{
$\hat{G}_n\Big(\big(h,v,z\big),\big(h',v,z\big)\Big)=\frac{1}{n}\sum_{i=1}^n\Big\{X_i\big(h,v,z\big)X_i\big(h',v,z\big)\Big\}$.
}
\label{eq:sample cov}
\end{equation}
and then $\hat{G}_H(h,h')$ found by marginalising over $v,z$.

Once $\hat{G}_H$, $\hat{G}_V$ and $\hat{G}_Z$ are computed using the above forms, we perform the segmentation for each direction of $H$, $V$ and $Z$ using the procedure as introduced in Section \ref{sec:optimal segmentation}. This requires choosing $\rho$ for each direction, so now we have to choose a 3-dimensional grid $A=(A_h\times A_v\times A_z)$, where $\rho_h\in A_h$, $\rho_v\in A_v$ and $\rho_z\in A_z$. Once we obtain boundary points for each direction of the coordinate, we create 3D segments in the following way. Assume that we obtain a sequence of boundary points in $H$-direction as $H^*=\left\{h_1^*,h_2^*,...,h_{L_h}\right\}$, in $V$-direction as $V^*=\left\{v_1^*,v_2^*,...,v_{L_v}\right\}$, and in $Z$-direction as $Z^*=\left\{z_1^*,z_2^*,...,z_{L_z}\right\}$. Then, 3D segments of $X$ are built as $X_{[l]}=\Big\{X(h,v,z): h\in(h_{j-1},h_j],v\in(v_{k-1},v_k],z\in(z_{q-1},z_q]\Big\}$, for all $j=1,...,L_h$, $k=1,...,L_v$ and $q=1,...,L_z$, where $l=1,...,L=L_h\times L_v\times L_z=$. The segments produced in this way will have a cuboid form. The brain has folded appearance and is round in shape. So the issue of approximation error can arise when trying to divide it into 3D cuboids. To avoid this issue, we set the size of each segment of the brain fairly small. As long as an element of brain images, e.g., voxel, displays high resolution, approximation error caused by separability would be minimal.

\subsection{Computational issues}
\label{sec:computational issues}
We save computational cost of the segmentation procedure via separability. Further, as addressed in Section \ref{sec:selection of segments}, we reduce the cost by adopting and modifying the idea as used in NOVAS \citep{Ferraty14}. Specifically, unlike NOVAS, where $q=p$ is assumed, we build potential subsets based on segments, so it would be more natural to set $q=L$, where $L$ is the total number of segments. In this way, only $\mathcal{O}(\sqrt{L}^2)=\mathcal{O}(L)$ number of potential subsets are explored in each step of the selection procedure. In our analysis of brain image data, however, the estimated $L$ is a few of thousands, and so $\mathcal{O}(L)$ is still quite large for the upper bound to the capability of computational resources. Instead, we use $q=L/2$ and save the cost even further. For instance, when $L=4000$, with $q=L$, the number of explored subsets at each step of the selection is 64, while with $q=L/2$, it is 45. The computational gain does not seem very large for single run of our subsampling scheme, however, as we repeat the selection procedure on subsamples 100 times, and each repetition involves $|B|$ number of estimation where $B$ is the considered set of tuning parameters, we can save $100\times|B|\times (64-45)$ number of computational operations.

Our method is computationally intensive. So we speed up computation by parallelization of the procedure. For instance, using a desktop with a 4-core, 3.4 GHz processor with 16GB RAM, the run time for the analysis of ADNI's FDG PET data presented in Section \ref{sec:3D applications} was less than 19 hours, where the total number of voxels involved was 1,408,000.

\section{3D numerical study}
\label{sec:study 3d}
We have conducted simulations with 3D functional data, but unlike the 1D case in the supplementary material, comparison with other methods is not made as they were developed only in the context of 1D functional data, and are not easily extended. We also apply the FuDoS methodology to the ADNI's PET brain image data.

\subsection{3D simulation}
\label{sec:sim2}
\begin{figure}
\centering
\includegraphics[width=.8\textwidth]{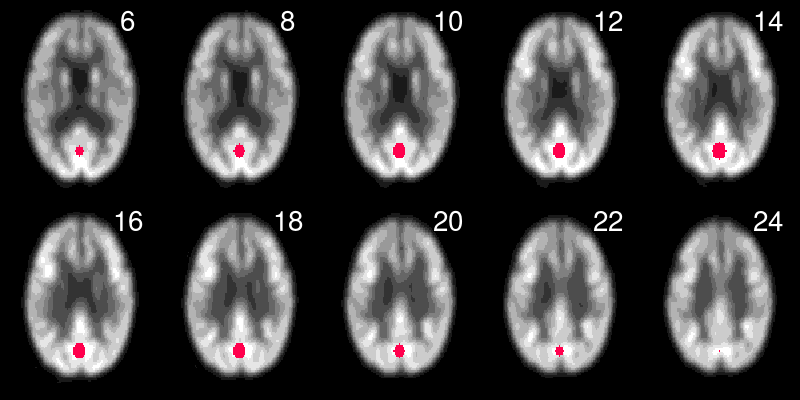}
\caption{True $\beta(h,v,z)$ overlaid on a randomly chosen PET brain image. The number on top is $z'=2(z-46)$, where $z$ is an axial slice number.}
\label{fig:true beta image}
\end{figure}

To realistically imitate brain images, we generate datasets based on ADNI's PET brain images. The details of acquisition and preprocessing of the images will be given in the supplementary material.
\begin{itemize}
\item The original ADNI's PET brain image, $X_i(h,v,z)$, $i=1,...,n_{t}$, with $n_t=1403$, lower index $t$ here means total, displays a grid of size $(160\times 160\times 96)$. To facilitate the computational time, we reduce the size to $(120\times 120\times 10)$, taking axial slices located at $z=51,...,60$ (focusing on central part of the brain) in the coordinate space, and eliminating some voxels outside of the brain.
\item We define the coefficient image as piecewise constant:
\[
    \beta(h,v,z)=
\begin{cases}
    10, & \text{if } (h-60)^2+(v-30)^2+(z-5)^2 \leq 5^2,\\
     0,   & \text{otherwise},
\end{cases}
\]
where $(h,v,z)$ means voxel location in the brain. Figure \ref{fig:true beta image} illustrates the true $\beta(h,v,z)$ overlaid on a randomly chosen individual's PET brain image. The number on top of each plot is $z'=2(z-48)$, where $z$ is an axial slice number of the brain.
\item Based on $X_i(h,v,z)$ and $\beta(h,v,z)$, we generate
\begin{equation*}
Y_i=\ip{X_i,\beta}+\epsilon_i, \quad i=1,...,n_t=1403,
\end{equation*}
where $\epsilon_i\sim N(0,\sigma^2)$, with $\sigma^2$ controlled by signal-to-noise ratio ($SNR$), i.e., $SNR=\text{var}(\tilde{Y}_i)/\sigma^2$ with $\tilde{Y}_i=\ip{X_i,\beta}$.
\end{itemize}

\begin{sidewaystable}
\centering
\resizebox{.9 \textwidth}{!}{%
\begin{tabular}{crrrrrrrrrr}
  \toprule
  & $\pi=0.05$ & 0.15 & 0.25 & 0.35 & 0.45 & 0.55 & 0.65 & 0.75 & 0.85 & 0.95\\
  \cmidrule{2-11}
 $n=\boldsymbol{1000},SNR=\boldsymbol{20}$\\
\textbf{RMSE (SE)}($\times 10^{-5}$)  & 6.46(2.29) & 8.2(1.66) & 6.68(1.71) & 5.62(1.79) & 4.88(1.89) & 4.44(2.06) & $\boldsymbol{4.17(2.38)}$ & 4.25(3.12) & 4.66(4.04) & 5.83(4.25) \\
$P_1\times 10^{2}$ & $\boldsymbol{100(0)}$ & $\boldsymbol{100(0.9)}$ & 99.9(3.1) & 99.5(6.1) & 98.2(8.8) & 94.9(12) & 91.3(15.6) & 84.3(18.7) & 71.9(18.9) & 55.3(15.8) \\
$P_2\times 10^{2}$ & 19.4(4) & 30.8(6.3) & 42(6.2) & 51.4(4.9) & 58.6(4.3) & 63.2(5.9) & $\boldsymbol{65.9(9.2)}$ & 65.3(13.4) & 61(15.5) & 50(14.2) \\
 \hline
 \\
 $n=\boldsymbol{200},SNR=\boldsymbol{20}$\\
\textbf{RMSE (SE)}($\times 10^{-5}$) & 8.66(2.29) & 8.15(1.66) & 6.96(1.71) & 6.1(1.79) & 5.6(1.89) & $\boldsymbol{5.39(2.06)}$ & 5.45(2.38) & 5.86(3.12) & 6.74(4.04) & 8.83(4.25) \\
$P_1\times 10^{2}$  & $\boldsymbol{100(0)}$ & 99.9(0.9) & 99.4(3.1) & 97.5(6.1) & 94.5(8.8) & 89.4(12) & 83.3(15.6) & 73.5(18.7) & 58.4(18.9) & 37(15.8) \\
$P_2\times 10^{2}$ & 18.5(4) & 32.8(6.3) & 45.2(6.2) & 53.9(4.9) & 59.8(4.3) & 62.6(5.9) & $\boldsymbol{63(9.2)}$ & 60(13.4) & 51.5(15.5) & 34.8(14.2) \\
 \hline
 \\
 $n=\boldsymbol{1000},SNR=\boldsymbol{10}$\\
\textbf{RMSE (SE)}($\times 10^{-5}$)  & 7.88(2.29) & 7.53(1.66) & 6.07(1.71) & 5.26(1.79) & 4.81(1.89) & 4.46(2.06) & $\boldsymbol{4.45(2.38)}$ & 4.6(3.12) & 5.43(4.04) & 6.75(4.25) \\
$P_1\times 10^{2}$  & $\boldsymbol{100(0)}$ & 99.9(0.9) & 99.5(3.1) & 98.4(6.1) & 95.7(8.8) & 91.9(12) & 87.5(15.6) & 78(18.7) & 64.4(18.9) & 46.1(15.8) \\
$P_2\times 10^{2}$& 22.4(4) & 37.5(6.3) & 48.5(6.2) & 56.3(4.9) & 61.5(4.3) & 64.6(5.9) & $\boldsymbol{65.5(9.2)}$ & 63.1(13.4) & 56.3(15.5) & 42.7(14.2) \\
 \hline
 \\
$n=\boldsymbol{200},SNR=\boldsymbol{10}$\\
\textbf{RMSE (SE)}($\times 10^{-5}$)  & 9.06(2.29) & 7.86(1.66) & 6.45(1.71) & 5.69(1.79) & $\boldsymbol{5.29(1.89)}$ & 5.32(2.06) & 5.61(2.38) & 6.39(3.12) & 7.57(4.04) & 11.95(4.25) \\
 $P_1\times 10^{2}$ & $\boldsymbol{100(0)}$ & 99.8(0.9) & 98.6(3.1) & 96.3(6.1) & 92.3(8.8) & 85.8(12) & 76.5(15.6) & 64(18.7) & 46.7(18.9) & 22.8(15.8) \\
$P_2\times 10^{2}$ & 19.8(4) & 36.9(6.3) & 49.4(6.2) & 57.1(4.9) & 61.8(4.3) & $\boldsymbol{63.2(5.9)}$ & 60.8(9.2) & 54.7(13.4) & 42.5(15.5) & 21.9(14.2) \\
 \hline
 \\
 $n=\boldsymbol{1000},SNR=\boldsymbol{5}$\\
\textbf{RMSE (SE)}($\times 10^{-5}$)  & 8.63(2.29) & 6.95(1.66) & 5.64(1.71) & 5.05(1.79) & 4.66(1.89) & $\boldsymbol{4.61(2.06)}$ & 4.77(2.38) & 5.28(3.12) & 6.16(4.04) & 8.71(4.25) \\
$P_1\times 10^{2}$ & $\boldsymbol{100(0)}$ & 99.9(0.9) & 99.5(3.1) & 97.5(6.1) & 93.6(8.8) & 89.4(12) & 82.4(15.6) & 70.7(18.7) & 55(18.9) & 34.6(15.8) \\
$P_2\times 10^{2}$ & 26.1(4) & 43(6.3) & 53.7(6.2) & 60.2(4.9) & 63.9(4.3) & $\boldsymbol{65.4(5.9)}$ & 64.5(9.2) & 59.8(13.4) & 49.5(15.5) & 32.7(14.2) \\
 \hline
 \\
 $n=\boldsymbol{200},SNR=\boldsymbol{5}$\\
\textbf{RMSE (SE)}($\times 10^{-5}$) & 10.32(2.29) & 8.17(1.66) & 6.75(1.71) & 6.15(1.79) & $\boldsymbol{6.1(1.89)}$ & 6.55(2.06) & 7.21(2.38) & 8.11(3.12) & 11.45(4.04) & 16.69(4.25) \\
$P_1\times 10^{2}$ & $\boldsymbol{100(0)}$ & 99.6(0.9) & 97.2(3.1) & 92.4(6.1) & 85.6(8.8) & 75.2(12) & 63(15.6) & 47.4(18.7) & 29.4(18.9) & 10.4(15.8) \\
$P_2\times 10^{2}$ & 20.6(4) & 40(6.3) & 52.9(6.2) & 59.7(4.9) & $\boldsymbol{61.7(4.3)}$ & 59(5.9) & 53.3(9.2) & 42.5(13.4) & 27.7(15.5) & 10.2(14.2) \\
 \hline
 \\
 $n=\boldsymbol{1000},SNR=\boldsymbol{2.5}$\\
\textbf{RMSE (SE)}($\times 10^{-2}$)  & 8.7(2.29) & 6.44(1.66) & 5.35(1.71) & 4.76(1.79) & $\boldsymbol{4.71(1.89)}$ & 4.98(2.06) & 5.45(2.38) & 6.33(3.12) & 7.88(4.04) & 11.89(4.25) \\
$P_1\times 10^{2}$ & $\boldsymbol{100(0)}$ & 99.7(0.9) & 98.2(3.1) & 94.2(6.1) & 89.1(8.8) & 81.7(12) & 71.2(15.6) & 56.6(18.7) & 39.7(18.9) & 20.3(15.8) \\
$P_2\times 10^{2}$ & 29.9(4) & 47.1(6.3) & 56.9(6.2) & 62(4.9) & $\boldsymbol{64.2(4.3)}$ & 63.1(5.9) & 58.8(9.2) & 49.9(13.4) & 36.9(15.5) & 19.6(14.2) \\
\hline
\\
 $n=\boldsymbol{200},SNR=\boldsymbol{2.5}$\\
\textbf{RMSE (SE)}($\times 10^{-5}$) & 11.8(2.29) & 8.74(1.66) & 7.23(1.71) & $\boldsymbol{6.88(1.79)}$ & 7.1(1.89) & 7.86(2.06) & 9.1(2.38) & 12.23(3.12) & 15.82(4.04) & 18.3(4.25) \\
$P_1\times 10^{2}$ & $\boldsymbol{100(0)}$ & 98.6(0.9) & 93.2(3.1) & 84.1(6.1) & 73.6(8.8) & 60.1(12) & 44.8(15.6) & 27.1(18.7) & 13.5(18.9) & 5.5(15.8) \\
$P_2\times 10^{2}$ & 21.5(4) & 42.9(6.3) & 54.6(6.2) & $\boldsymbol{58.5(4.9)}$ & 57(4.3) & 50.5(5.9) & 40.1(9.2) & 25.5(13.4) & 13.1(15.5) & 5.4(14.2) \\
 \bottomrule
\end{tabular}}
\caption{Simulation results of 3D case. The results include mean of RMSE, $P_1$ and $P_2$ and their standard deviation (in brackets). Minimum of RMSE and maximum of $P_1$ and $P_2$ in each case of simulation are written in bold.}
\label{tab:3D Tab}
\end{sidewaystable}

\begin{figure}
\centering
		 \subfigure{\includegraphics[scale=0.4]{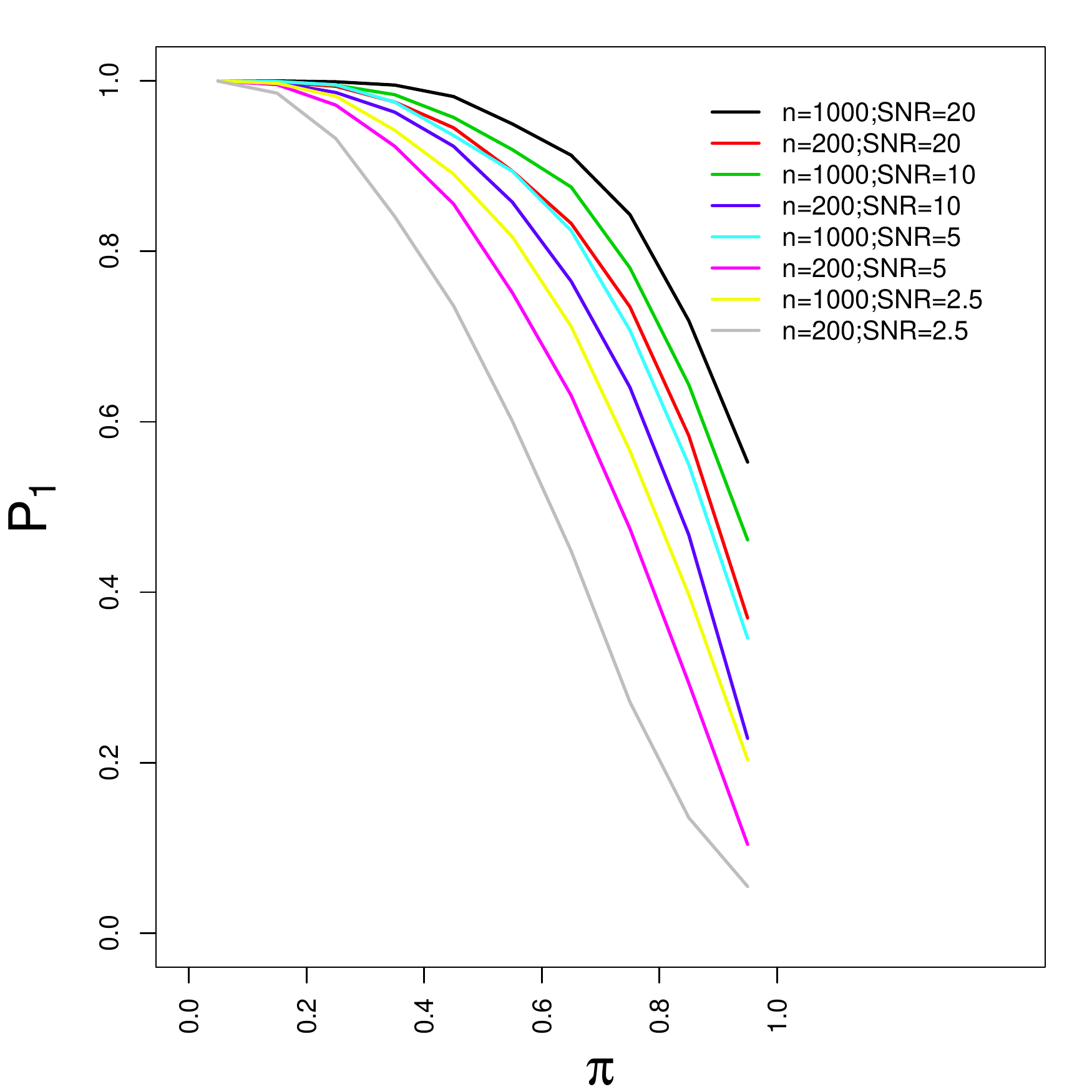}}
  	\subfigure{\includegraphics[scale=0.4]{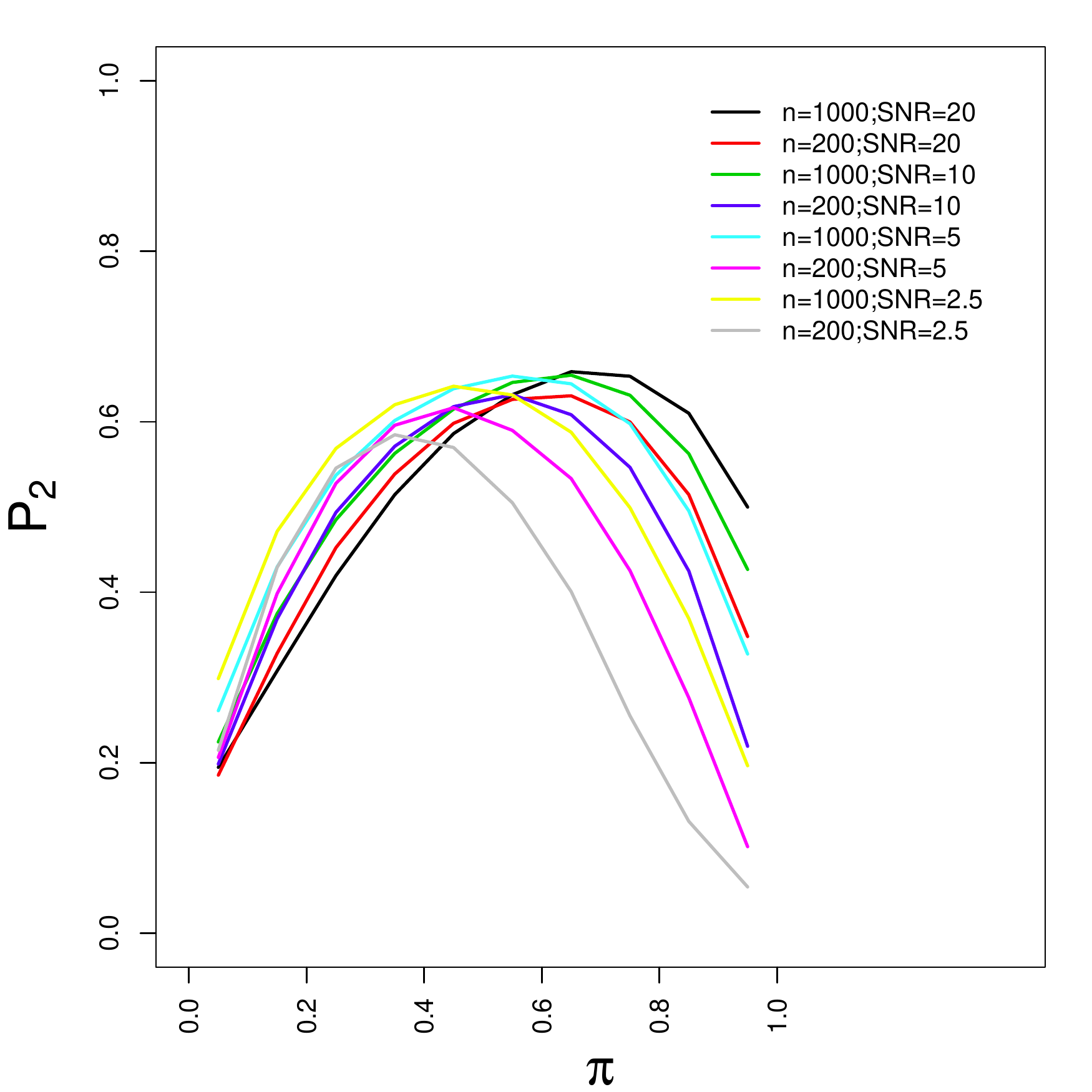}}
  \caption{3D simulation result: Mean of $P_1$ and $P_2$ calculated with different values of $\pi$.}
\label{fig:3D sim path fig}
\end{figure}

\begin{figure}
\centering
	  \subfigure[$n=1000,SNR=20$]{\includegraphics[width=.45\textwidth]{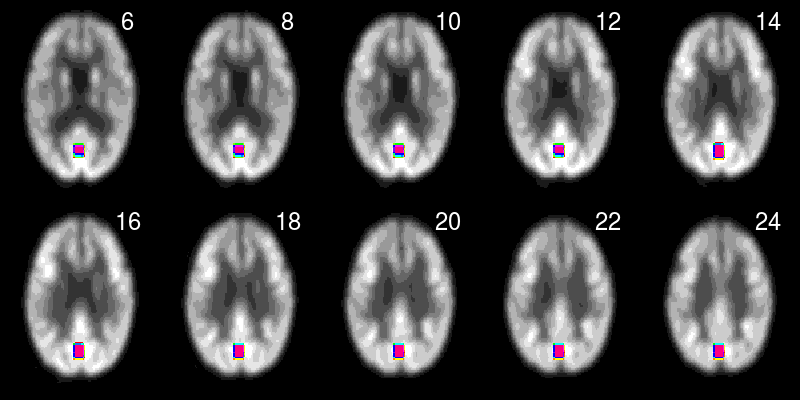}
	  \label{fig:3D.case1.coef}}
  	\subfigure[$n=200,SNR=20$]{\includegraphics[width=.45\textwidth]{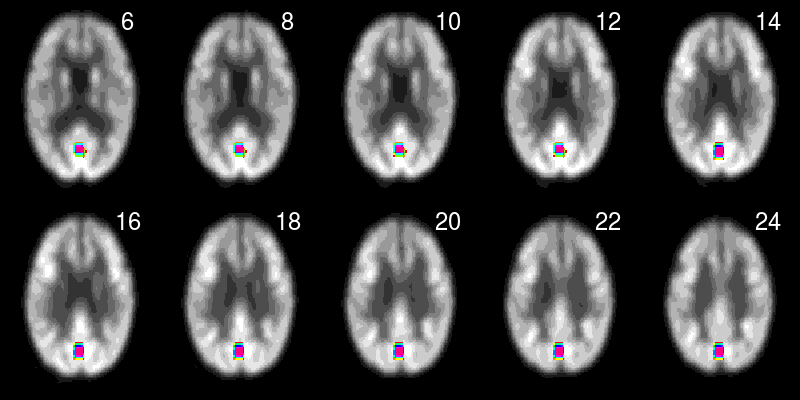}
	  \label{fig:3D.case2.coef}}
	
	   \subfigure[$n=1000,SNR=10$]{\includegraphics[width=.45\textwidth]{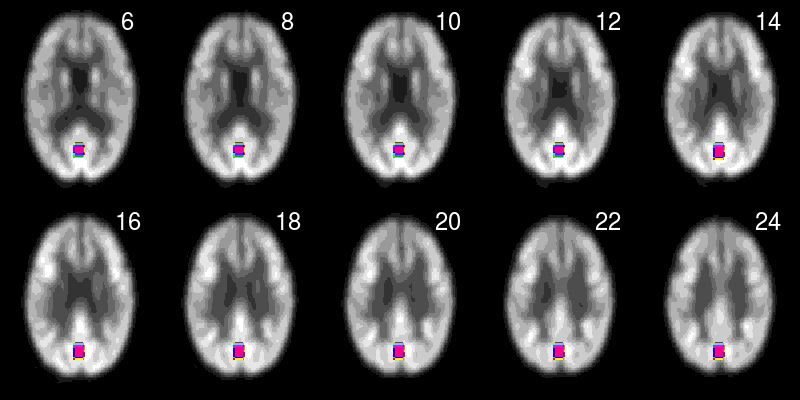}
	  \label{fig:3D.case3.coef}}
  	\subfigure[$n=200,SNR=10$]{\includegraphics[width=.45\textwidth]{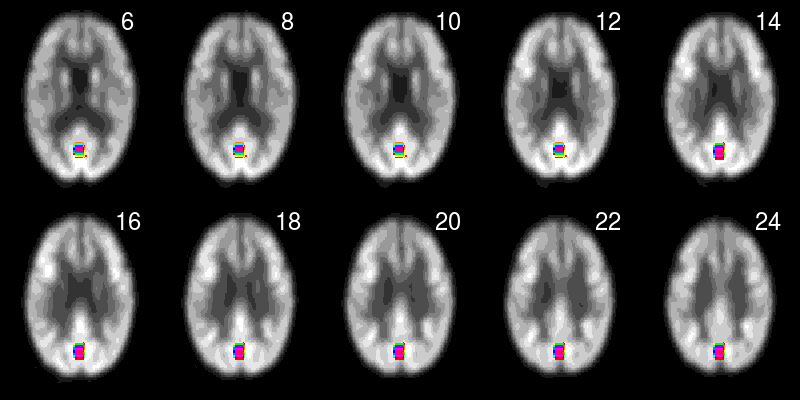}
	  \label{fig:3D.case4.coef}}
	
	   \subfigure[$n=1000,SNR=5$]{\includegraphics[width=.45\textwidth]{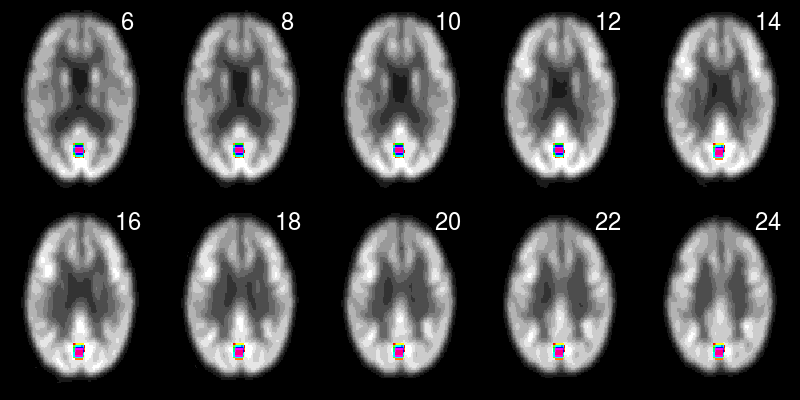}
	  \label{fig:3D.case5.coef}}
  	\subfigure[$n=200,SNR=5$]{\includegraphics[width=.45\textwidth]{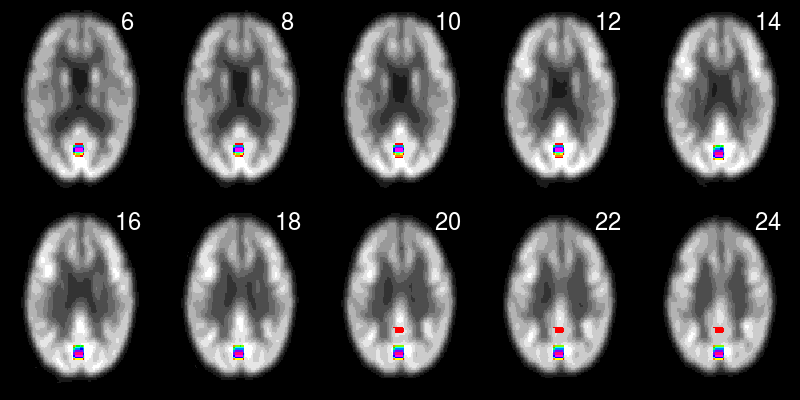}
	  \label{fig:3D.case6.coef}}
	
	  \subfigure[$n=1000,SNR=2.5$]{\includegraphics[width=.45\textwidth]{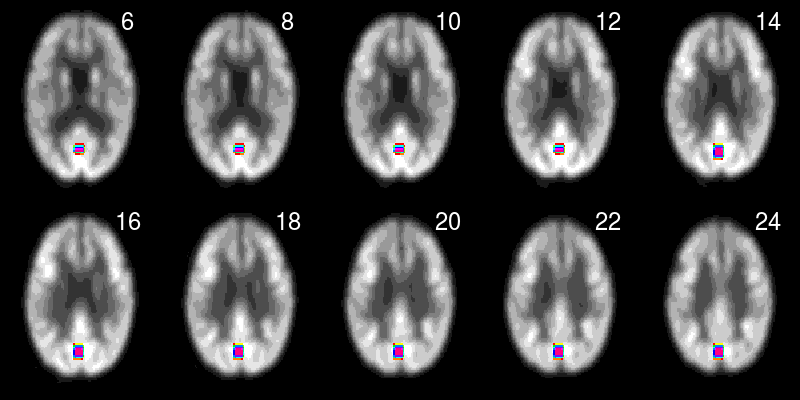}
	  \label{fig:3D.case7.coef}}
	  \subfigure[$n=200,SNR=2.5$]{\includegraphics[width=.45\textwidth]{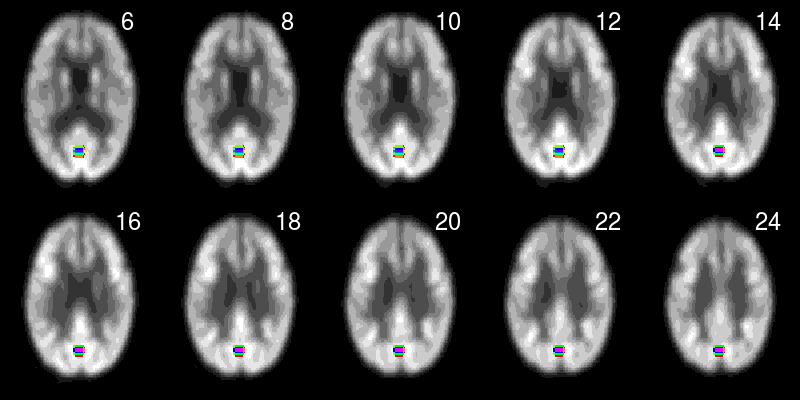}
	  \label{fig:3D.case8.coef}}
	   \subfigure{\includegraphics[scale=0.5]{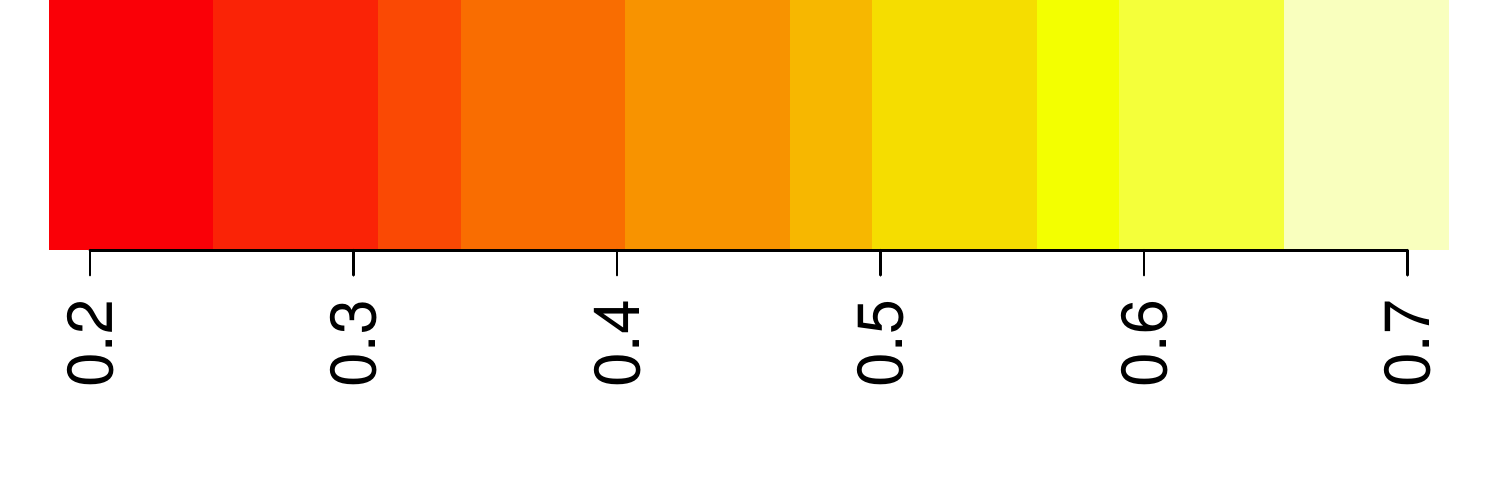}}
	   \caption{Selected stable subregions with selection probability higher than $0.4$ in each case of simulation setting. The number on top is $z'=2(z-46)$, where $z$ is axial slice number. Colour indicates estimated maximum selection probability.}
  	\label{fig:3D sim coef fig}
\end{figure}

We consider whether the FuDoS methodology is able to identify true subdomain of $X$ that have an association with $Y$ under various levels of error and sample size. To address this, we consider four values of $SNR=\left\{2.5,5,10,20\right\}$ and two values of $n=\left\{200,1000\right\}$, producing eight simulations in total. In each case, $n$ brain images (either 200 or 1000) are randomly drawn from the total $n_t=1403$ brain images, and $Y_i$ are generated using the above form with different values of $SNR$. For a random subsample of the data with size $\lceil n/2 \rceil$, we select the most predictive subset for each pair of $(\rho,c)\in B$, returning $X^{{\rho,c}}_{\mathcal{J}^*}$. In choosing $B=(A\times C)$, where $A=(A_h\times A_v\times A_z)$, $\rho_h\in A_h$, $\rho_v\in A_v$, $\rho_z\in A_z$ and $c\in C$, we employ $A_h=A_v=A_z=\left\{0.01, 0.03\right\}$ and $C=\left\{0.01\right\}$, so in total 8 pairs of $(\rho,c)$ are involved. The subsampling procedure is repeated 100 times to estimate the selection probability as in (\ref{equ:cutoff}), and the stable subdomain, see (\ref{equ:selection proba}), is determined when the maximum selection exceeds the cut-off $\pi$. For fitting $\beta_{[l]}$, see (\ref{eq:local linear model}), a piecewise constant basis is used.

We evaluate the selection performance using $P_1$ and $P_2$ as follows. Denote the true segment, where $\beta(t)\neq 0$, by $\mathcal{X}^*$, and write the estimated stable subdomain by $\hat{\mathcal{X}}_{stable}^{\pi}$. Denoting $|\mathcal{A}|$ the size of any set $\mathcal{A}$, we compute
\begin{equation}
P_1=\frac{|\mathcal{X}^*\cap \hat{\mathcal{X}}_{stable}^\pi|}{|\mathcal{X}^*|}.
\label{eq:P1}
\end{equation}
The size of $\hat{\mathcal{X}}_{stable}^\pi$ tends to get larger as $\pi$ decreases, so that the selected stable subregion would include the true set with high probability as $\pi$ becomes smaller. When trying to recover the true set $\mathcal{X}^*$, a natural goal would be to include as few false segments as possible. To penalize the rate of falsely identified subregion, we also measure
\begin{equation}
P_2=\frac{|\mathcal{X}^*\cap \hat{\mathcal{X}}_{stable}^\pi|}{|\mathcal{X}^*\cup \hat{\mathcal{X}}_{stable}^\pi|}.
\label{eq:P2}
\end{equation}

The predictive ability of each method is also investigated. We sample $n_v=200$ pairs of test observations, where the lower index $v$ means \textit{validation}, except for the samples included in the training set, carry out prediction and calculate the root mean squared error (RMSE) on the test samples, defined as $(\sum_{i=1}^{n_v}(Y_i-\hat{Y_i})^2/n_v)^{1/2}$, where $\hat{Y_i}$ is the predicted value of $i$th test sample, $Y_i$. As explained in Section \ref{sec:constructing predictive model}, we build a predictive model based on $\mathcal{X}_{stable}^\pi$ for each $\pi$, so prediction RMSE is measured for each $\pi$.

Each simulation is iterated 100 times and the average and the standard deviation of $P_1$, $P_2$ and RMSE are reported in Table \ref{tab:3D Tab}. The minimum of RMSE and the maximum of $P_1$ and $P_2$ in each case of simulations are written in bold. When $n$ and $SNR$ get smaller, it becomes more difficult to indentify the true set with high probability. Therefore, one should select a small value for $\pi$ to guarantee that the selected stable subregion encompasses the true set. Indeed, the best range of $\pi$ achieving fairly small values of RMSE is $0.35<\pi<0.65$, and it tends to decrease as $n$ and $SNR$ become smaller. In the supplementary material, it is seen  that the optimal range of $\pi$ in the 1D functional case was higher, i.e., it was $0.55<\pi<0.85$. This is not surprising as the problem of $n<<p$ is more severe in case of 3D functional data, i.e., $p=128$ and $n=(50,800)$ versus $p=144,000$ and $n=(200,1000)$. The average of $P_1$ and $P_2$ for different values of $\pi$ is plotted in Figure \ref{fig:3D sim path fig}. The value of $P_1$ approaches to 1 as $\pi$ approaches to 0. Both $n$ and $SNR$ have an impact on the selection result, but the effect of $n$ seems to be stronger. An estimated stable subdomain with $\pi=0.4$ for a simulation in each case of simulation setting is shown in Figure \ref{fig:3D sim coef fig}. The colour indicates maximum value of selection probability.

\subsection{Analysis of ADNI's FDG PET}
\label{sec:3D applications}
The ADNI's PET data used in this analysis consists of $n=1302$ individuals, including participants from all of the ADNI's study phases: 402 individuals from ADNI-1, 127 from ADNI-GO and 773 from ADNI-2. Recall that a total number of available ADNI's PET brain images used in our 3D simulation study was $n_t=1403$, as it included brain images from the same subject acquired at different time of visit. More detailed demographic features of the involved individuals are summarized in Table \ref{tab:summary demo}. Acquisition and preprocessing parameters of the data set are explained in the supplementary material. 

\begin{table}
\caption{\label{tab:summary demo}Demographic features of ADNI's individuals used in the analysis.}
\centering
\begin{tabular}{lrrrrr}
 \toprule
 & Total \# & Female \# & Age & Education & MMSE \\
  \cmidrule{2-6}
Normal & 288 & 137 & 74.43(5.91) & 16.31(2.79) & 28.99(1.21) \\
  MCI & 773 & 337 & 72.77(7.3) & 16.14(2.7) & 27.97(1.73) \\
  AD & 241 & 99 & 75.13(7.89) & 15.34(2.95) & 23.21(2.12) \\
   \bottomrule
\end{tabular}
\end{table}

Using the preprocessed ADNI's PET brain images, the goal is to identify subregions of the brain associated with cognitive deficit, and we use the subjects' mini-mental state examination (MMSE) scores as a measure of cognitive ability. Typically, the range of the MMSE is $0-30$, and it tends to decline as AD progresses as seen in Table \ref{tab:summary demo}. Assuming that only relatively few areas of the brain are truly related to cognitive ability, we applied the proposed methodology to the data set. In some situations, brain images may have predictive power for a clinical outcome as the images are related to one or more demographic characteristics, as a form of confounder, that drive the relation. Investigating the presence of confounding effects would be useful in practice because scalar covariates are generally much simpler than images to acquire \citep{Reiss15}. To test whether the subjects' demographic variables have an association with cognitive decline, the MMSE was linearly regressed on gender, age and years of education. The achieved R-square was 0.06, so these scalar covariates seem to be unrelated. In comparison, the predictive R-square achieved by the brain images using the proposed method is larger than 0.3, as discussed later in the section, therefore, we decided not to adjust MMSE scores for demographic variables. To reduce memory burden and facilitate computational time, we decrease the size of the brain to $(120\times 120\times 55)$, taking axial slices located at $z=16,...,70$, and eliminating voxels outside of the brain. When we fitted the same model involving the complete brain slices using participants in the ADNI 1 study phrase only (in this case $n=402$), clusters of voxels in these discarded brain slices appeared to be irrelevant, so we dropped the seemingly redundant brain slices from the analysis. The size of the original PET brain image data with $n=1302$ subjects was approximately 13 GB, but after the reduction, the size has decreased to 8.2 GB.
\begin{table}
\caption{\label{tab:n.stable.voxels}Total number of voxels in selected stable subregions for each value of $\pi$.}
\centering
\resizebox{.9 \textwidth}{!}{%
\begin{tabular}{crrrrrrrrrrrrr}
 \toprule
$\pi$ & 0.1 & 0.15 & 0.2 & 0.25 & 0.3 & 0.35 & 0.4 & 0.45 & 0.5 & 0.55 & 0.6 & 0.65 & 0.7 \\
 \cmidrule{2-14}
$|\mathcal{X}_{stable}^\pi|$ & 6842 & 4267 & 2770 & 1960 & 1466 & 1192& 921 & 779 & 560 & 533 & 399 & 119 & 0 \\
 \bottomrule
   \end{tabular}}
\end{table}

\begin{figure}
\centering
\subfigure{\includegraphics[width=.5\textwidth]{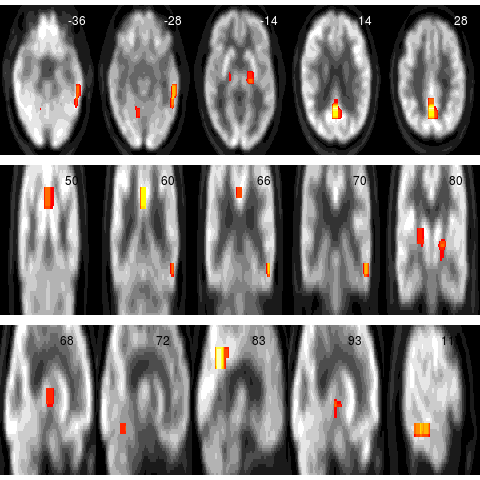}}
\subfigure{\includegraphics[width=.5\textwidth]{adni_clour-eps-converted-to.pdf}}
\caption{Selected subregions of the ADNI's PET brain associated with cognitive decline with selection probability higher than $0.2$. Colour indicates estimated maximum selection probability. The first row shows axial view of the brain, the second is sagittal and the third is coronal.}
  		\label{fig:stab.prob.adni}
  		
\end{figure}
First, we identify subregions of the brain, associated with the MMSE scores, selected by the FuDoS method with high probability. For each random subsample of the data set with size $\lceil n/2 \rceil$, $n=1302$, we fit FuDoS to the subsamples, gaining $X^{{\rho,c}}_{\mathcal{J}^*}$, for each $(\rho,c)\in B$, where $B=(A\times C)$, with $A=(A_h\times A_v\times A_z)$, $A_h=A_v=A_z=\left\{0.01,0.03\right\}$ and $C=\left\{0.01\right\}$. The size of each segment is set to be $3^3-7^3$ cubes, and with the considered values in $A$, the estimated number of segments $\hat{L}$ is approximately ranges from 2100 to 7500. When the full brain was used, the same segmentation rule led to approximately $5100<\hat{L}<11,100$ . Note that the total number of voxels involved in the analysis is 792,000.

For fitting $\beta_{[l]}$, piecewise constant basis is used. Thus, a total number of regression parameters to be estimated is the same as the number of segments in the set ${\mathcal{J}}$, except an intercept. Because ADNI's PET data is high resolution, i.e., 1.5 mm of voxel size, and the size of each segment is quite small, i.e., each segment includes relatively small number of voxels, using a constant form for $\beta_{[l]}$ is not be too restrictive. In other words, influences of neighbouring voxels in the same segment are likely to be similar. The above subsampling procedure is repeated 100 times, and for each $(\rho,c)$, the probability of each voxel being included in $X^{{\rho,c}}_{\mathcal{J}^*}$ is estimated. We take maximum of the probability over $B$, and determine stable subregions $\mathcal{X}_{stable}^\pi$ for each $\pi$ as in (\ref{equ:cutoff}). Table \ref{tab:n.stable.voxels} reports the number of selected stable voxels for different values of $\pi$.

The brain is composed of three parts: the brainstem, cerebellum and cerebrum (the largest part of the brain), with the surface of the cerebrum called the cortex. The cortex has a folded appearance, and each fold (gyrus) a groove between a sulcus. Cerebral cortex contains most of the brain's neuronal cell bodies (grey matter), and includes regions of the brain involved in sensory perception such as seeing and hearing, memory, emotions, speech, decision making. In the analysis of AD, grey matter is most of interest as it has been clinically proven that measures of semantic (fact-based) and short-term memory have a significant positive correlation with grey matter volume in older people \citep{Resnick03, Buckner04}. In contrast, no association was found between white matter volume and variability in cognitive functions. The cerebrum is divided into four lobes: frontal, parietal, temporal and occipital, and each lobe is composed of several areas. Each of the brain areas serves specific functions. They do not function alone, but often, work together having complex relationships with each other.

In AD patients, there is an overall shrinkage of brain tissue. The sulci are noticeably widened, and big shrinkage in the gyri is often found. Moreover, in many brain imaging studies of early AD, including \citet{gusnard01searching}, decreased metabolism has been found predominantly in the posterior cingulate cortex (associated with yellow clusters in Figure \ref{fig:stab.prob.adni}), medial temporal lobe (associated with orange/red clusters in Figure \ref{fig:stab.prob.adni}) and inferior parietal lobe.

Figure \ref{fig:stab.prob.adni} presents the estimated selection probability for the selected stable subregions. To aid visualization, we view the brain in three different anatomical planes: axial plane (top row), coronal plane (middle row) and sagittal plane (bottom row). We note that the clusters of voxels identified in our analysis agrees well with the two expected anatomical brain regions. Firstly, the big yellow/orange clusters, in axial planes of 14 and 28, in coronal planes of $50-66$, and in sagittal plane of 83, are associated with the posterior cingulate cortex. The posterior cingulate cortex (PCC) is the posterior part of the cingulate cortex, situated in the upper part of the limbic lobe, surrounded by the precuneus and the retrosplenial cortex. The PCC is known to have memory-related functions, and many previous studies have found abnormal patterns of the brain in this region in AD patients \citep{Foster84, Minoshima95, Minoshima97, Huang02cingulate}. For example, \citet{Minoshima95} proposed a fully automated approach to discriminating probable AD patients from normal control (NC) subjects, and a statistically significant reduction in glucose metabolism was found in potential AD patients in most of cortical areas, including the parietal, temporal and frontal cortex. The profound brain abnormalities in AD patients in the posterior parietal lobes were also demonstrated in \citet{Foster84}, see their Fig 1. Similar to our analysis, a regression approach was used in earlier studies to reveal the brain abnormalities. \citet{Minoshima97} linearly regressed the MMSE on each voxel of the brain in cortical areas, and the results indicate a marked metabolic reduction in the PCC in patients who are at the very early stage of AD, see their Fig 1 and 2.

Functional brain imaging techniques also have been widely used for the analysis for AD as AD is closely related to the changes in the functional connectivity among different brain regions \citep{Fransson08}. Based on functional MRI (fMRI) and diffusion tensor inmaging (DTI) data, \citet{Zhou08abnormal} investigated the functional connectivity maps of representative of NC, mild cognitive impairment (MCI) and early AD subjects, and claimed a significant reduction of fiber bundles in the PCC in the groups of early AD and MCI, compared with the NC group, see their Fig 1 and 2. Moreover, according to \citet{Huang02cingulate}, a reduction in relative blood flow of the posterior cingulate gyrus could be found, at least two years before the patients are clinically diagnosed as AD.

The second expected brain region identified from our analysis is hippocampus and medial temporal lobe, related to the red clusters of voxels, in axial locations between -36 and -14, in coronal locations between 60 and 80, and in sagittal locations between 68 and 72 and 93 and 116. Many studies have identified the anatomical components of the brain system that govern memory function in the medial temporal lobe, and this neural system consists of the hippocampus and adjacent, including entorhinal, perirhinal, and parahippocampal cortex \citep{Squire91}. Considering the well-known medial temporal lobe memory system, it is not surprising that early symptoms of AD are associated with pathological change and loss of neurons in this lobe \citep{Jobst94, Jack97,Visser99,Dickerson08}.

\begin{figure}
\centering
\subfigure{\includegraphics[width=.5\textwidth]{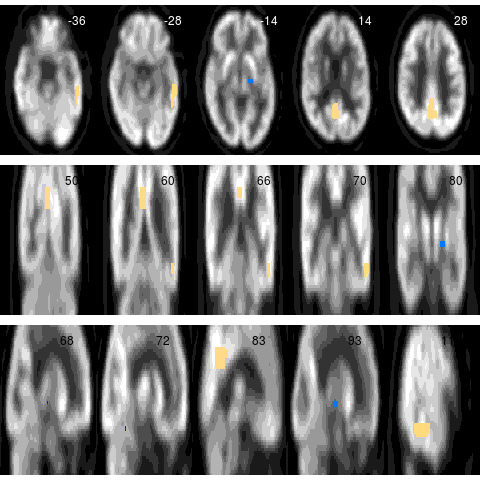}}
\subfigure{\includegraphics[width=.5\textwidth]{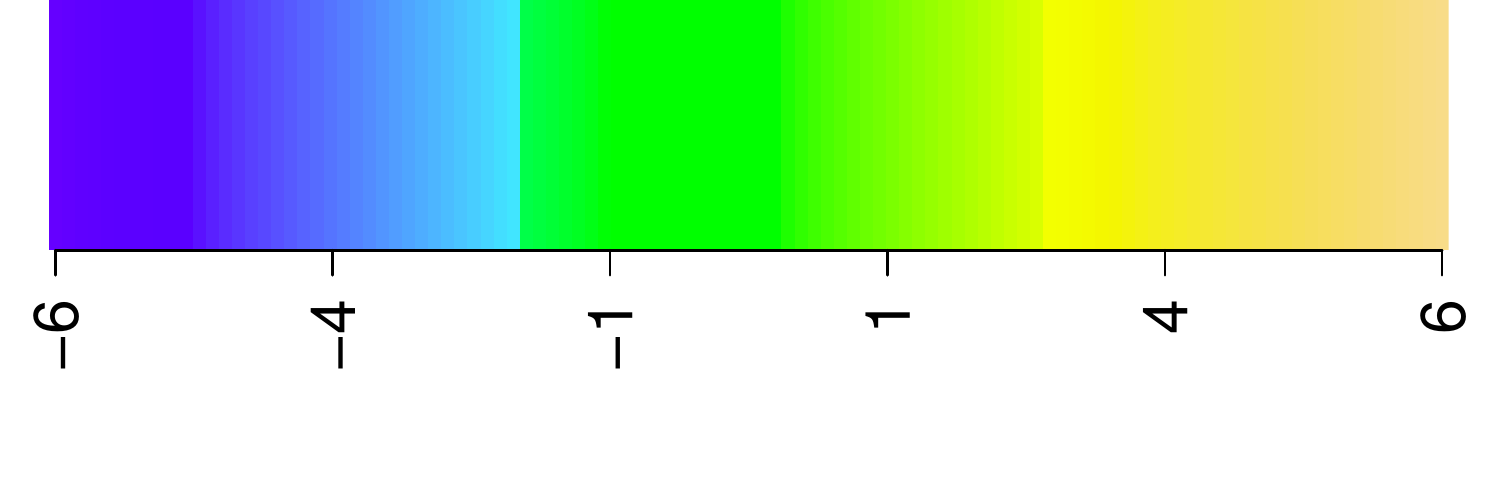}}
\caption{Mean effects of selected subregions over 100 bootstrap samples.}
  		\label{fig:bootstrap.mean.adni}

  		\end{figure}

We now investigate mean effects of the selected stable subregions on the MMSE scores. In this analysis, the used cut-off value is $\pi=0.3$, which results in a total number of selected voxels of 1466, as given in Table \ref{tab:n.stable.voxels}. We used a density based clustering algorithm for spatial data \citep{Ester96, Sander98} to group the selected voxels as explained in Section \ref{sec:constructing predictive model}, resulting in four subregions (four clusters of voxels). To obtain the sampling distributions of the mean effects, bootstrapping is used. The obtained sampling distributions from 100 iterated bootstrapping are reported in Table \ref{tab:bootstrap dist}, and the mean of each effect is displayed in Figure \ref{fig:bootstrap.mean.adni}. The lower bound of the bootstrap-based confidence intervals of the subregion 1 and 3, associated with the voxels in hippocampus is negative. This might indicate that glucose consumption of the brain cells in hippocampal area has a negative relation with the cognitive ability, and the effects are statistically significant. While, the mean effects on the cognitive ability at the subregion 2 and 4, related to voxels in the PCC and some parts of the medial temporal lobe, is positive and statistically significant.

\begin{table}
\caption{\label{tab:bootstrap dist}Quantiles of mean effects over 100 bootstrap samples for selected stable subregions.}
\centering
\begin{tabular}{lrrrr}
 \toprule
 & subregion 1 & subregion 2 & subregion 3 & subregion 4\\
 \cmidrule{2-5}
2.5\% & -8.36 & 3.38 & -5.44 & 3.66 \\
  97.5\% & -3.56 & 6.00 & -1.65 & 6.74 \\
 \bottomrule
\end{tabular}
\end{table}

\begin{table}
\caption{\label{tab:prediction results}Average number of voxels in selected stable subregions, and predictive R-square and RMSE for each value of $\pi$.}
\centering
\resizebox{.9 \textwidth}{!}{%
\begin{tabular}{lrrrrrrrrrrrrr}
  \toprule
$\pi$ & 0.2 & 0.225 & 0.25 & 0.275 & 0.3 & 0.325 & 0.35 & 0.375 & 0.4 & 0.425 & 0.45 & 0.475 & 0.5 \\
 \cmidrule{2-14}
size & 2625 & 2274 & 1878 & 1628 & 1357 & 1205 & 1044 & 951 & 844 & 773 & 675 & 629 & 563 \\
  R-square & 0.35 & 0.35 & 0.35 & 0.34 & 0.34 & 0.30 & 0.30 & 0.27 & 0.25 & 0.24 & 0.23 & 0.23 & 0.22 \\
  RMSE & 2.12 & 2.12 & 2.13 & 2.14 & 2.14 & 2.20 & 2.21 & 2.25 & 2.28 & 2.29 & 2.31 & 2.31 & 2.33 \\
  \bottomrule
\end{tabular}}
\end{table}

Finally, we exploit a 10-fold cross validation to perform prediction. Similar to the analysis of gasoline data in the supplementary material, we leave out 10\% of observations as a validation set, use the rest to train the model, including the identification of the stable subregions through 100 repeated subsampling, and perform the prediction on the data points that have been left out. Repeating this procedure for each sample fold, we aggregate the predicted values, and compute the predictive RMSE and predictive R-square, defined as $1-\sum(Y_i-\hat{Y}_{-j,i})^2/\sum(Y_i-\bar{Y})^2$, where $\hat{Y}_{-j,i}$ is the predictive value of $Y_i$, based on the model fitted from training samples except samples in $j$th sample fold. As introduced in Section \ref{sec:constructing predictive model}, predictive models, denoted by $M_{\pi}$, were built for each value of $\pi$, and the results of each predictive model are reported in Table \ref{tab:prediction results}. The achieved predictive R-square with $\pi<0.3$ is 0.35. We note that \citet{Wang14} also carried out 10-fold cross validated prediction of MMSE scores based on the ADNI's PET brain images, including $n=403$ number of subjects, and the produced predictive R-square using their method was 0.26.

\section{Conclusion}
\label{sec:conclusion}
In this article, we have introduced a domain selection method which, in the context of functional data analysis, effectively identifies subregions of the brain associated with a clinical outcome of interest. The methodology is general, so it can be applied to any kind of data, where a predictor $X$ is functional and a response $Y$ is scalar. The methodology is composed of two stages of estimation. We first segment $X$ into several small parts based on the correlation structure. Then, potential subsets are built using the obtained segments and their predictive performance are evaluated to select the best subset. To account for functional features of $X$, two functional regression approaches, either penalized splines, or piecewise constant basis, are considered for fitting the regression function. We used a subsampling scheme, i.e., stability selection criterion to stabilize the selected subset and we found that this selection scheme provides several advantages to the proposed method. It increases the selection probability at boundaries of the true segments. Moreover, it reduces the rate of falsely selected subregion. The proposed method also has a practical advantage. Due to the induced sparseness, the results naturally provide more interpretable information about the relations between the regions and the outcome. We also investigated predictive ability of selected stable subregions. Our 1-D numerical results given in the supplementary material suggest that the selected stable sets can be used for building predictive models, and they can outperform competing methods in prediction for a range of $\pi$.

One possible criticism of the proposed method would be that the current segmentation procedure does not account for the response $Y$. So, one can imagine to obtain a more relevant segmentation using a conditional covariance function of $X$ given $Y$, or using an approximated covariance function based on functional partial least squares decomposition. However, due to the high-dimensional nature of the brain image data, the extension is not trivial, and at least the numerical cases considered in this article may not suggest the need of a more complex approach for the segmentation as the current methodology can select the true subset with high probability. Another possible criticism would be that functional linear regression can be too restrictive in some situations, and so one can expect to achieve better predictive performance by replacing the functional linear model with more flexible models, such as multiple functional index models, functional projection pursuit models, or pure nonparametric models. However, these flexible models are computationally very intensive, therefore they might be practically impossible to apply to the brain image data. Also, at least, in the analysis considered in this article, a linearity assumption on the regression function does not seem to be restrictive.

\bibliography{mybib1}
\appendix
\section{Some simulation results of segmentation}
\begin{figure}
\centering
\subfigure{\includegraphics[width=.22\textwidth]{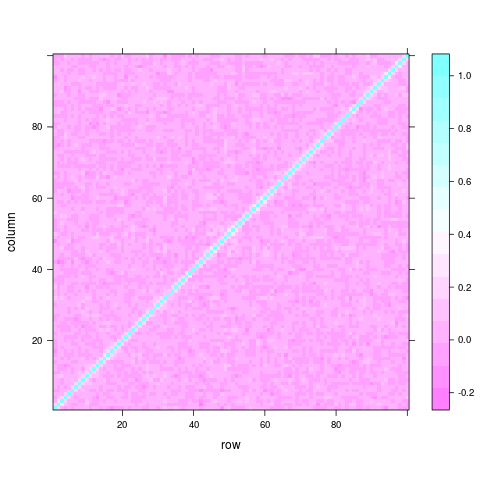}}
\subfigure{\includegraphics[width=.22\textwidth]{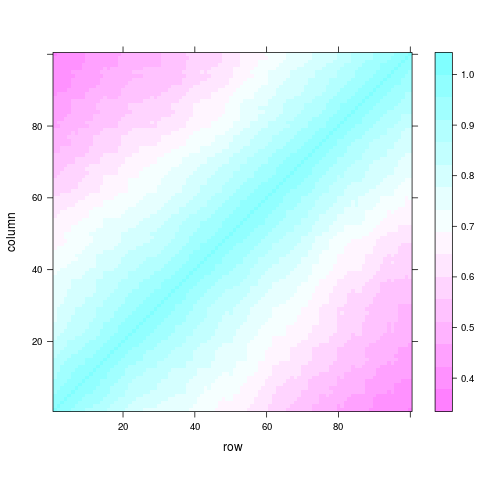}}
\subfigure{\includegraphics[width=.22\textwidth]{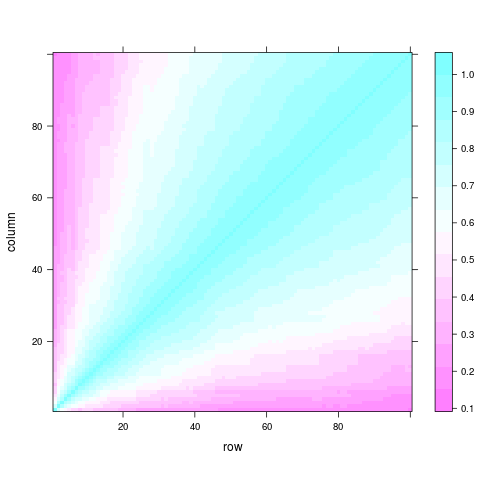}}
\subfigure{\includegraphics[width=.22\textwidth]{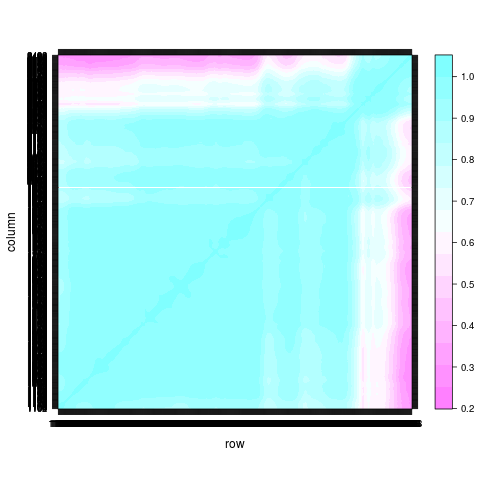}}
		
\subfigure{\includegraphics[width=.22\textwidth]{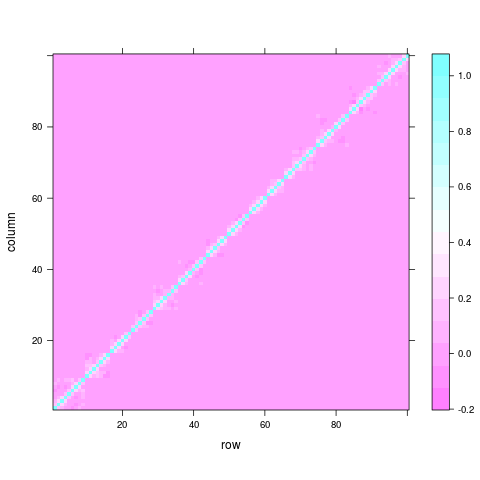}}
\subfigure{\includegraphics[width=.22\textwidth]{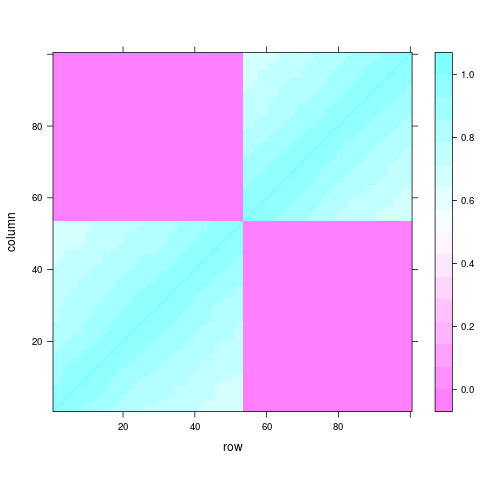}}
\subfigure{\includegraphics[width=.22\textwidth]{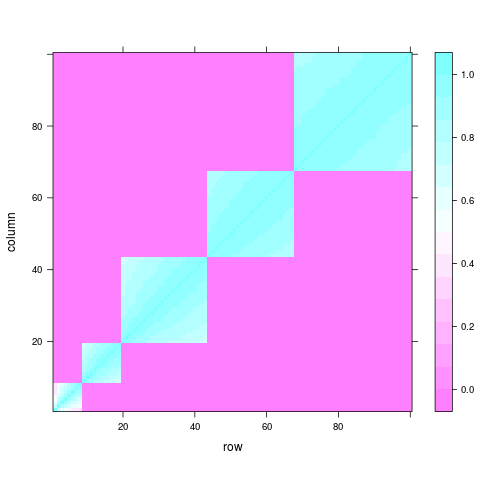}}
\subfigure{\includegraphics[width=.22\textwidth]{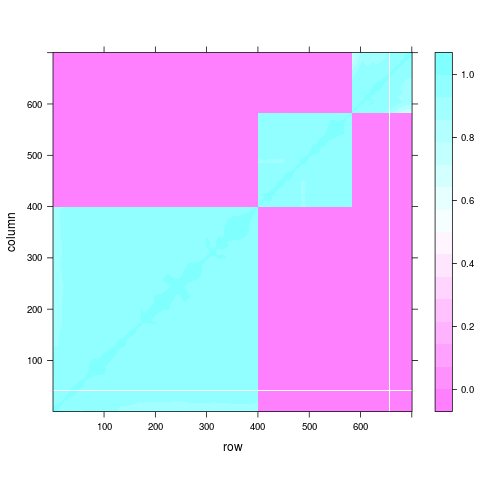}}
\caption{Simulated segmentation results. First row shows simulated correlation matrices and the estimated segmentation results are given below. First and second column are simulated from AR(1) process with AR coefficients of 0.2 and 0.99, respectively. Third column is simulated from Brownian motion and fourth column is associated with biscuit dough data \citep{R.package.fds}.}
\label{fig:ex segmentation}
\end{figure}

Figure \ref{fig:ex segmentation} displays some simulated segmentation results. The first row shows simulated correlation matrices and the corresponding estimated segmentation results are given below. The algorithm seems to perform reasonably well as it splits region with low correlation, while keeping region with high correlation intact.

\section{Alzheimer's Disease Neuroimaing Initiative (ADNI)}
\label{sec:adni}
Having been launched in October 2004 and now in its third phase, Alzheimer's Disease Neuroimaing Initiative (ADNI) is a global initiative, which unites researchers with reliable data to understand prevention, progression, as well as treatment of MCI and early AD. The ADNI data are extensive, including MRI and PET brain images, genetics, cerebrospinal fluid (CSF) and blood biomarkers, as well as clinical and neuropsychological tests, and are available to the general scientific community. Coinvestigators of ADNI are wide-ranging, involving the National Institutes of Health, the Food and Drug Administration, academic institutions, nonprofit groups and private sectors, such as drug and medical-imaging companies. Initial participants in ADNI have been recruited from over 50 sites across North America with an age range of 55-90, approximately 200 cognitively normal controls, 400 with MCI and 200 with early AD. New participants were recruited during each phase of the study, and they are followed and reassessed over time.

\section{Acquisition and preprocessing of FDG PET images}
\label{sec:preprocessing}
Before scanning, the tracer FDG (fludeoxyglucoser), an analogue of glucose is injected through a vein, and the tracer travels through blood and collects in brain tissues. The injected dose of FDG was $5.0\pm0.5 mCi$, and the subjects were scanned from 30 to 60 minutes post-injection. This procedure generates either six five-minute frames (for ADNI-1), or four five-minute frames (for ADNI-GO and ADNI-2). All subject underwent neurological examinations within three months of the scanning, involving the memory, effective memory and mini-mental state examination (MMSE). The MMSE invented by \citet{Folstein75} examines various cognitive abilities, including orientation to time and place, immediate and delayed recall of three words, attention and calculation, language and visuo-constructural functions. It is often used by clinicians alongside patients' medical history, symptoms, physical exams and the results of other tests, including brain images to diagnose dementia and assess progression and severity of the disease. Typically, the range of the MMSE is $0-30$, and it tends to decline as AD progresses as seen in Table \ref{tab:summary demo}. The PET brain image scans were preprocessed by the following steps. Each frame was coregistered to the first frame of the raw image file. Six or four coregistered frames were averaged to create a single PET image. Each subject's coregistered and averaged PET image from the baseline PET scan was reoriented into a standard grid $(160\times160\times96)$, displaying cubic voxels of size 1.5 mm, and the anterior-posterior axis of each subject is parallel to a AP-CP line. Finally, each image set is filtered with a scanner-specific filter function to produce images of a uniform isotropic resolution of 8 mm FWHM. The detailed description of the PET preprocessing is described in \url{http://adni.loni.usc.edu/methods/pet-analysis/pre-processing/}.

\section{1D numerical study}
\label{sec:study 1d}
As a proof of concept, in addition to analyzing three dimensional data, we investigate the properties on 1-D functional data. This allows comparison with other 1-D methods in the literature. We assess the performance of the proposed methodology on simulated 1D functional data, and compare the results with FLiRTI \citep{James09} and WLasso \citep{Zhao12}. We also demonstrate the methods on the gasoline data set \citep{Rpackage:refund}.
\subsection{1D simulation}
\label{sec:sim1}
Two forms of $X(t)$ are considered and the details of the simulation settings are following.
\begin{itemize}
\item Case 1: $X_i(t)$, $i=1,...,n$, $t\in[0,1]$, are generated from $ARMA(2,2)$ and discretized at equi-spaced points:
\begin{equation}
\resizebox{.9 \textwidth}{!}
{
$X_i(t_j)=0.8X_i(t_{j-1})-0.1X_i(t_{j-1})+e_i(t_j)-0.1e_i(t_{j-1})+0.9e_i(t_{j-2}),\quad j=1,...,128,$
}
\label{sim:ARMA}
\end{equation}
where $e_i(t_j)\sim N(0,1)$.

\item Case 2: $X_i(t)$, $i=1,...,n$, $t\in[0,1]$, are a linear combination of cubic B-splines with interior knots placed at $1/16,...,15/16$ and coefficients, that is,
\begin{equation}
X_i(t)=\sum_j^{16}c_{ij}\phi_j(t),
\label{sim:splines}
\end{equation}
where $c_{ij}\sim N(0,4)$, and $\phi_j(t)$ are B-spline basis functions.

\item In both cases we use a disconnected smooth function for $\beta(t)$ with flat region. Specifically,
\[
    \beta(t)=
\begin{cases}
    0.5\cos(40t-\pi)+2t,& \text{if } 0.39\leq t < 0.44,\\
    0.5\sin(40t-\pi)+2t,& \text{if } 0.73\leq t < 0.79,\\
    0,              & \text{otherwise},
\end{cases}
\]
where the range of $t$ is given as approximated values. Because of the discretization, each segment can be expressed as a set of consecutive design point $t_j$ with an index $j$ as
\[
    \beta(t_j)=
\begin{cases}
    0.5\cos(40t_j-\pi)+2t_j,& \text{if } 50\leq j\leq 56,\\
    0.5\sin(40t_j-\pi)+2t_j,& \text{if }  94\leq  j\leq 100,\\
    0,              & \text{otherwise}.
\end{cases}
\]
\item Based on $X_i(t)$ and $\beta(t)$, we simulate
\begin{equation*}
Y_i=<X_i,\beta>+\epsilon_i,\quad i=1,...n,
\label{sim:Y}
\end{equation*}
where $\epsilon_i\sim N(0,\sigma^2)$ is an observational noise, with $\sigma^2$ determined by signal-to-noise ratio ($SNR$), i.e., $SNR=\text{var}(\tilde{Y}_i)/\sigma^2$ with $\tilde{Y}_i=\ip{X_i,\beta}$. In approximation, we can write $\tilde{Y}_i=\ip{X_i,\beta}\approx \frac{1}{128}\sum_{j=1}^{128}X_i(t_j)\beta(t_j)$.
\end{itemize}

To examine the performance of the proposed method under various simulation settings, we vary the sample size as $n=\left\{50, 800\right\}$, and the size of noise as $SNR=\left\{2.5,5,10,20\right\}$, producing eight simulations for each case of $X$. To determine the stable subdomains, we randomly draw subsamples of size $\lceil n/2 \rceil$ a hundred times, returning 100 selected subsets $X^{{\rho,c}}_{\mathcal{J}^*}$, for each pair of $(\rho,c)\in B$. Then, we estimate the selection probability of each subdomain $\mathcal X$ as in (\ref{equ:selection proba}), and take its maximum over $B$. When choosing $B=(A\times C)$ with $\rho\in A$ and $c\in C$,  we consider a single set for $C$ as $C=\left\{0.01\right\}$, and consider $A=\left\{.02, .035, .04, .05, .06\right\}$ in case 1 of $X$, and $A=\left\{0, .03, .04, .06, .08\right\}$ in case 2 of $X$. The set $A$ is chosen to achieve two goals: 1) the size of each segment is such that it encompasses at least 5 equispaced $t_j$'s and 20 at most; and 2) different values of $\rho$ result in different levels of segmentation $\hat{L}$, where $\hat{L}$ is the estimated number of segments. In case 1 of $X$, $A=\left\{.02, .035, .04, .05, .06\right\}$ approximately results in $\hat{L}=\left\{19, 17, 15, 12, 8\right\}$ on average, and in case 2 of $X$, $A=\left\{0, .03, .04, .06, .08\right\}$ leads to $\hat{L}=\left\{18, 14, 12, 10, 8\right\}$. For fitting $\beta_{[l]}$ as in (\ref{eq:local linear model}), penalized B-splines \citep{Marks99, Cardot03}, with the smoothing parameter selected by generalized cross validation \citep{Craven79} is used. To fit FLiRTI \citep{James09}, one must choose three values of tuning parameters: 1) the penalty parameter to adjust the level of sparsity; 2) the weight to be placed on the zeroth derivative relative to the higher order derivative; and 3) the derivative order to assume sparsity in. We use the default choice of the zeroth and the second derivative order to impose sparsity, and select the penalty and weight parameter using a 5-fold CV, which is the default setting of the R-code provided by the authors. WLasso \citep{Zhao12} involves three tuning parameters: 1) the number of coefficients retained for prediction; 2) the penalty parameter associated with the regularizing term; and 3) the min-scale adjusting the coarseness level of the wavelet decomposition, and the optimal values of these three parameters are chosen by a 5-fold CV. We evaluate the selection performance using $P_1$ as in (\ref{eq:P1}) and $P_2$ as in (\ref{eq:P2}). The prediction ability of FuDoS is also investigated. The predictive ability of each method is also investigated. We generate 1000 pairs of test samples, carry out prediction and calculate the RMSE on test samples. As explained in Section \ref{sec:constructing predictive model}, we build a predictive model based on $\mathcal{X}_{stable}^\pi$ for each $\pi$, so prediction RMSE is measured for each $\pi$.

\begin{sidewaystable}
\caption{Simulation results of 1D/case 1 of $X$. The results include mean of RMSE, $P_1$ and $P_2$ and their standard deviation (in brackets). Minimum of RMSE and maximum of $P_1$ and $P_2$ in each case of simulation are highlighted in bold.}
\label{tab:1D Tab1}
\centering
\resizebox{.9 \textwidth}{!}{%
\begin{tabular}{crrrrrrrrrrr}\toprule
&\multicolumn{8}{c}{\textbf{FuDoS}} & \multicolumn{1}{c}{\textbf{FLiRTI}}& \multicolumn{1}{c}{\textbf{WLasso}}\\
\cmidrule{2-9}
&$\pi=\boldsymbol{0.25}$&$\boldsymbol{0.35}$&$\boldsymbol{0.45}$&$\boldsymbol{0.55}$&$\boldsymbol{0.65}$&$\boldsymbol{0.75}$&$\boldsymbol{0.85}$&$\boldsymbol{0.95}$&&\\
\midrule
$n=\boldsymbol{800},SNR=\boldsymbol{20}$\\
\textbf{RMSE (SE)}($\times 10^{-2}$)  & 1.13(0.09) & 1.11(0.09) & 1.09(0.09) & 1.07(0.09) & 1.05(0.11) & 1.02(0.12) & 0.98(0.16) & $\boldsymbol{0.88(0.25)}$ & 1.26(0.16) & 23.95(0.61) \\
$P_1\times 10^{2}$ & $\boldsymbol{100(0)}$ & $\boldsymbol{100(0)}$ & $\boldsymbol{100(0)}$ & $\boldsymbol{100(0)}$ & $\boldsymbol{100(0)}$ & $\boldsymbol{100(0)}$ & 99.71(1.41) & 98.64(2.82) &
$\boldsymbol{100(0)}$ &  \\
$P_2\times 10^{2}$ & 26.4(3.44) & 29.32(3.93) & 32.5(4.84) & 36.35(6) & 41.46(6.97) & 48.13(8.12) & 58.17(9.56) & $\boldsymbol{75.37(10.8)}$ & 38.12(31.94) &  \\
\hline
\\
$n=\boldsymbol{50},SNR=\boldsymbol{20}$\\
\textbf{RMSE (SE)}($\times 10^{-2}$) & 4.22(1) & 3.8(0.9) & 3.56(0.85) & 3.26(0.83) & 3.02(0.84) & $\boldsymbol{3.01(0.91)}$ & 4.35(2.71) & 9.01(3.74) & 4.64(1.11) & 24.44(2.38) \\
$P_1\times 10^{2}$ & $\boldsymbol{100(0)}$ & 99.86(1.01) & 99.79(1.22) & 99.36(2.51) & 97.86(4.37) & 93.14(9.08) & 79.36(16.99) & 49.43(14.37) & 96.14(6.29) &  \\
$P_2\times 10^{2}$ & 21.28(4.16) & 28.3(5.67) & 33.62(5.79) & 40.6(6.64) & 49.56(8.41) & 57.95(9.28) & $\boldsymbol{61.31(13.66)}$ & 46.04(12.26) & 31.68(22.52) &  \\
\hline
\\
$n=\boldsymbol{800},SNR=\boldsymbol{10}$\\
\textbf{RMSE (SE)}($\times 10^{-2}$)  & 1.33(0.15) & 1.29(0.15) & 1.27(0.15) & 1.22(0.15) & 1.2(0.15) & 1.17(0.16) & $\boldsymbol{1.14(0.24)}$ & 1.17(0.39) & 1.58(0.27) & 23.9(0.63) \\
$P_1\times 10^{2}$& $\boldsymbol{100(0)}$ & $\boldsymbol{100(0)}$ & $\boldsymbol{100(0)}$ & $\boldsymbol{100(0)}$ & 99.93(0.71) & 99.71(1.41) & 99.07(2.62) & 96.07(4.8) & $\boldsymbol{100(0)}$ &  \\
$P_2\times 10^{2}$  & 26.99(3.69) & 30.65(4.75) & 34.2(5.72) & 38.37(6.83) & 44.36(8.12) & 50.68(9.17) & 60(9.98) & $\boldsymbol{74.95(11.04)}$ & 69.63(25.4) &  \\
\hline
\\
$n=\boldsymbol{50},SNR=\boldsymbol{10}$\\
\textbf{RMSE (SE)}($\times 10^{-2}$)  & 5.9(1.35) & 5.16(1.22) & 4.7(1.21) & 4.2(1.18) & 3.93(1.04) & $\boldsymbol{3.9(1.16)}$ & 4.71(2.04) & 8.32(3.41) & 6.17(1.75) & 24.47(2.83) \\
$P_1\times 10^{2}$ & $\boldsymbol{100(0)}$ & $\boldsymbol{100(0)}$ & 99.64(1.87) & 98.43(3.74) & 95.93(6.28) & 89.36(10.18) & 76.29(15.14) & 49.35(13) & 92.79(9.28) &  \\
$P_2\times 10^{2}$  & 20.36(3.58) & 27.84(5.19) & 33.86(6.03) & 41.1(6.73) & 50.22(7.97) & 57.95(9.14) & $\boldsymbol{60.58(12.04)}$ & 46.62(12.39) & 39(24.3) &  \\
\hline
\\
$n=\boldsymbol{800},SNR=\boldsymbol{5}$\\
\textbf{RMSE (SE)}($\times 10^{-2}$)  & 1.7(0.25) & 1.62(0.25) & 1.59(0.25) & 1.52(0.24) & 1.49(0.25) & 1.48(0.31) & $\boldsymbol{1.47(0.31)}$ & 1.97(0.55) & 1.74(0.3) & 23.65(3.14) \\
$P_1\times 10^{2}$& $\boldsymbol{100(0)}$ & $\boldsymbol{100(0)}$ & $\boldsymbol{100(0)}$ & $\boldsymbol{100(0)}$ & 99.71(1.41) & 98.93(2.76) & 97.57(3.69) & 88.36(7.84) & 99.71(1.41) &  \\
$P_2\times 10^{2}$  & 27.87(3.91) & 31.99(4.97) & 35.69(5.66) & 40.09(6.22) & 46.19(8.23) & 53.32(8.61) & 63.5(10.08) & $\boldsymbol{73.86(8.55)}$ & 52.69(15.67) &  \\
\hline
\\
$n=\boldsymbol{50},SNR=\boldsymbol{5}$\\
\textbf{RMSE (SE)}($\times 10^{-2}$) & 7.98(2.02) & 6.43(1.81) & 5.69(1.42) & $\boldsymbol{4.89(1.22)}$ & 5.02(1.74) & 5.77(2.74) & 7.89(3.61) & 11.03(3.21) & 7.06(2.33) & 24.99(2.53) \\
$P_1\times 10^{2}$ & $\boldsymbol{100(0)}$ & 99.86(1.43) & 99.36(2.29) & 97.36(5.71) & 89.86(11.85) & 79.43(17.15) & 60.98(18.38) & 44.77(13.55) & 92.29(11.38) &  \\
$P_2\times 10^{2}$  & 20.9(4.51) & 29.94(6.5) & 37.07(7.98) & 45.44(8.89) & 53.56(9.29) & $\boldsymbol{57.03(12.59)}$ & 51.35(13.67) & 41.22(11.43) & 36.94(15.38) &  \\
\hline
\\
$n=\boldsymbol{800},SNR=\boldsymbol{2.5}$\\
\textbf{RMSE (SE)}($\times 10^{-2}$)  & 2.16(0.34) & 2.08(0.33) & 2.01(0.34) & 1.97(0.32) & 1.95(0.34) & $\boldsymbol{1.9(0.39)}$ & 2.05(0.5) & 2.7(0.66) & 2.22(0.39) & 23.87(2.24) \\
$P_1\times 10^{2}$ & $\boldsymbol{100(0)}$ & $\boldsymbol{100(0)}$ & $\boldsymbol{100(0)}$ & $\boldsymbol{100(0)}$ & 99.29(2.38) & 97.93(4.1) & 93.5(6.43) & 82.43(8.09) & 99.64(1.56) &  \\
$P_2\times 10^{2}$  & 28.85(3.94) & 32.73(4.56) & 36.71(5.72) & 41.79(6.8) & 48.67(7.44) & 55.84(9.08) & 63.64(8.29) & $\boldsymbol{73(8.45)}$ & 36.28(10.98) &  \\
\hline
\\
$n=\boldsymbol{50},SNR=\boldsymbol{2.5}$\\
\textbf{RMSE (SE)}($\times 10^{-2}$)  & 9.79(2.22) & 7.85(2.14) & 7.27(2.5) & $\boldsymbol{7.24(2.78)}$ & 8.39(3.53) & 10.17(3.76) & 12.08(2.92) & 12.93(0.75) & 8.09(2.82) & 24.44(3.8) \\
$P_1\times 10^{2}$ &  $\boldsymbol{100(0)}$ & 98.5(6.83) & 93.43(13.57) & 86.43(17.3) & 72.51(19.66) & 58.22(18.66) & 44.49(11.92) & 38.31(7.34) & 91(12.38) &  \\
$P_2\times 10^{2}$  & 22.59(5.26) & 34.49(7.72) & 41.64(9.41) & 47.79(11.5) & $\boldsymbol{50.49(14.24)}$ & 47.74(15) & 40.18(11.76) & 36.68(7.76) & 26.01(7.59) &  \\
\bottomrule
\end{tabular}}
\end{sidewaystable}

\begin{sidewaystable}
\caption{Simulation results of 1D/case 2 of $X$. The results include mean of RMSE, $P_1$ and $P_2$ and their standard deviation (in brackets). Minimum of RMSE and maximum of $P_1$ and $P_2$ in each case of simulation are highlighted in bold.}
\label{tab:1D Tab2}
\centering
\resizebox{.9 \textwidth}{!}{%
\begin{tabular}{crrrrrrrrrrr}\toprule
&\multicolumn{8}{c}{\textbf{FuDoS}} & \multicolumn{1}{c}{\textbf{FLiRTI}}& \multicolumn{1}{c}{\textbf{WLasso}}\\
\cmidrule{2-9}
&$\pi=\boldsymbol{0.25}$&$\boldsymbol{0.35}$&$\boldsymbol{0.45}$&$\boldsymbol{0.55}$&$\boldsymbol{0.65}$&$\boldsymbol{0.75}$&$\boldsymbol{0.85}$&$\boldsymbol{0.95}$&&\\
\midrule
$n=\boldsymbol{800},SNR=\boldsymbol{20}$\\
\textbf{RMSE (SE)}($\times 10^{-2}$)  & 0.38(0.1) & 0.37(0.1) & 0.36(0.1) & 0.34(0.1) & 0.33(0.09) & $\boldsymbol{0.33(0.1)}$ & 0.35(0.1) & 1.77(2.95) & 0.44(0.1) & 16.25(0.71) \\
$P_1\times 10^{2}$ & $\boldsymbol{99.93(0.71)}$ & 99.64(1.87) & 99.14(2.74) & 97.64(3.94) & 93.93(5.87) & 88(8.18) & 77.93(11.13) & 59.38(13.33) & 98.64(3.32) &  \\
$P_2\times 10^{2}$ & 36.01(3.69) & 40.39(4.02) & 43.7(4.18) & 47.83(4.78) & 53.38(7.61) & 57.34(9.78) & $\boldsymbol{59.07(11.05)}$ & 51.88(10.63) & 32.04(7.18) &  \\
\hline
\\
$n=\boldsymbol{50},SNR=\boldsymbol{20}$\\
\textbf{RMSE (SE)}($\times 10^{-2}$) & 1.98(0.58) & 1.77(0.5) & 1.71(0.49) & 1.61(0.47) & $\boldsymbol{1.51(0.49)}$ & 2.46(2.55) & 5.74(3.73) & 7.4(2.65) & 1.86(0.56) & 16.11(1.48) \\
$P_1\times 10^{2}$ & $\boldsymbol{100(0)}$ & $\boldsymbol{100(0)}$ & 99.86(1.43) & 99.36(3.06) & 95.43(8.53) & 78.64(19.85) & 55.84(17.68) & 50.6(11.17) & 98.79(3.38) &  \\
$P_2\times 10^{2}$ & 20.46(2.81) & 27.02(3.68) & 32.17(3.83) & 39.07(5.09) & 49(8.63) & $\boldsymbol{52.57(13.84)}$ & 46.47(10.82) & 47.61(10.95) & 27.66(8.4) &  \\
\hline
\\
$n=\boldsymbol{800},SNR=\boldsymbol{10}$\\
\textbf{RMSE (SE)}($\times 10^{-2}$) & 0.39(0.11) & 0.38(0.11) & 0.37(0.11) & 0.36(0.11) & $\boldsymbol{0.35(0.11)}$ & 0.36(0.1) & 0.37(0.1) & 1.64(2.83) & 0.44(0.11) & 16.42(1.22) \\
$P_1\times 10^{2}$ & $\boldsymbol{99.86(1.01)}$ & 99.57(1.7) & 98.79(3.22) & 97.29(4.52) & 93.71(6.83) & 88.57(9.58) & 79.86(11.77) & 60.43(14.66) & 97.86(4.37) &  \\
$P_2\times 10^{2}$& 37.22(3.93) & 41.23(4.16) & 44.02(4.23) & 48.15(5.22) & 53.09(8.32) & 56.96(9.91) & $\boldsymbol{59.18(11.01)}$ & 52.64(11.48) & 31.79(7.02) &    \\
\hline
\\
$n=\boldsymbol{50},SNR=\boldsymbol{10}$\\
\textbf{RMSE (SE)}($\times 10^{-2}$)  & 1.97(0.55) & 1.81(0.47) & 1.73(0.44) & $\boldsymbol{1.65(0.42)}$ & 1.7(1.09) & 3.19(3.11) & 6.86(3.29) & 8.8(0.5) & 1.84(0.6) & 16.37(2.63) \\
$P_1\times 10^{2}$ & $\boldsymbol{100(0)}$ & $\boldsymbol{100(0)}$ & 99.86(1.01) & 99.29(3.13) & 93.29(10.97) & 74.43(19.82) & 53.21(14.27) & 43.88(6.43) & 98.43(3.3) &  \\
$P_2\times 10^{2}$  & 21.16(2.64) & 27.24(3.82) & 32.07(4.5) & 38.6(5.32) & 48.08(8.55) & $\boldsymbol{51.51(12.18)}$ & 46.37(11.7) & 41.35(5.9) & 28.33(8.26) &  \\
\hline
\\
$n=\boldsymbol{800},SNR=\boldsymbol{5}$\\
\textbf{RMSE (SE)}($\times 10^{-2}$) & 0.39(0.08) & 0.37(0.08) & 0.37(0.08) & 0.36(0.09) & 0.35(0.09) & $\boldsymbol{0.35(0.09)}$ & 0.39(0.11) & 1.91(3.13) & 0.45(0.11) & 16.25(0.81) \\
$P_1\times 10^{2}$& $\boldsymbol{100(0)}$ & 99.71(1.41) & 98.79(3.38) & 97.86(4.25) & 95.29(5.94) & 90.21(8.1) & 80.57(11.4) & 62.1(15.53) & 97.79(5.25) &  \\
$P_2\times 10^{2}$  & 36.57(3.51) & 40.64(3.95) & 43.86(3.9) & 48.17(5.04) & 53.87(8.55) & 58.45(10.68) & $\boldsymbol{60.39(11.21)}$ & 54.01(13.52) & 33.41(10.56) &  \\
\hline
\\
$n=\boldsymbol{50},SNR=\boldsymbol{5}$\\
\textbf{RMSE (SE)}($\times 10^{-2}$) & 1.96(0.56) & 1.86(0.48) & 1.72(0.46) & $\boldsymbol{1.67(0.45)}$ & 1.74(1.27) & 3.84(3.51) & 7.03(3.22) & 8.54(0.2) & 1.86(0.61) & 16.16(1.83) \\
$P_1\times 10^{2}$ & $\boldsymbol{100(0)}$ & $\boldsymbol{100(0)}$ & $\boldsymbol{100(0)}$ & 99(3.8) & 90.07(13.69) & 71.43(21.08) & 52.35(15.02) & 44.16(5.36) & 98.64(2.99) &  \\
$P_2\times 10^{2}$ & 20.93(2.45) & 26.98(3.19) & 32.5(3.98) & 39.28(4.75) & 45.87(8.01) & $\boldsymbol{48.84(13.82)}$ & 44.1(10.69) & 41.83(5.43) & 28.72(10) &  \\
\hline
\\
$n=\boldsymbol{800},SNR=\boldsymbol{2.5}$\\
\textbf{RMSE (SE)}($\times 10^{-2}$)  & 0.39(0.09) & 0.38(0.1) & 0.37(0.09) & 0.36(0.09) & $\boldsymbol{0.35(0.09)}$ & 0.36(0.09) & 0.38(0.11) & 1.52(2.69) & 0.44(0.09) & 16.26(0.69) \\
$P_1\times 10^{2}$ & $\boldsymbol{99.93(0.71)}$ & 99.64(1.87) & 98.57(3.52) & 96.64(5.32) & 93.64(7.1) & 88.14(8.63) & 79.07(10.71) & 59.6(14.33) & 98.21(4.47) &  \\
$P_2\times 10^{2}$ &36.21(4.08) & 40.83(4.83) & 44.81(5.27) & 48.92(5.84) & 55.25(9.43) & 59.16(11.63) & $\boldsymbol{61.19(12.29)}$ & 53(12.46) & 34.49(8.14) &  \\
\hline
\\
$n=\boldsymbol{50},SNR=\boldsymbol{2.5}$\\
\textbf{RMSE (SE)}($\times 10^{-2}$) & 1.93(0.49) & 1.81(0.46) & 1.7(0.44) & $\boldsymbol{1.62(0.42)}$ & 1.75(1.33) & 3.23(3.2) & 6.32(3.59) & 9.23(3.02) & 1.78(0.5) & 16.14(1.53) \\
$P_1\times 10^{2}$ & $\boldsymbol{100(0)}$ & $\boldsymbol{100(0)}$ & 99.93(0.71) & 99.21(3.03) & 91.21(13.04) & 74.78(18.81) & 54.7(15.68) & 39.29(7.77) & 98.14(4.14) &  \\
$P_2\times 10^{2}$ & 21.35(2.7) & 27.49(3.53) & 32.99(4.33) & 39.36(5.13) & 46.88(8.11) & $\boldsymbol{49.62(12.5)}$ & 44.7(10.67) & 36.27(8.25) & 29.73(8.65) &  \\
\bottomrule
\end{tabular}}
\end{sidewaystable}

\begin{figure}
\centering
	  \subfigure{\includegraphics[scale=0.4]{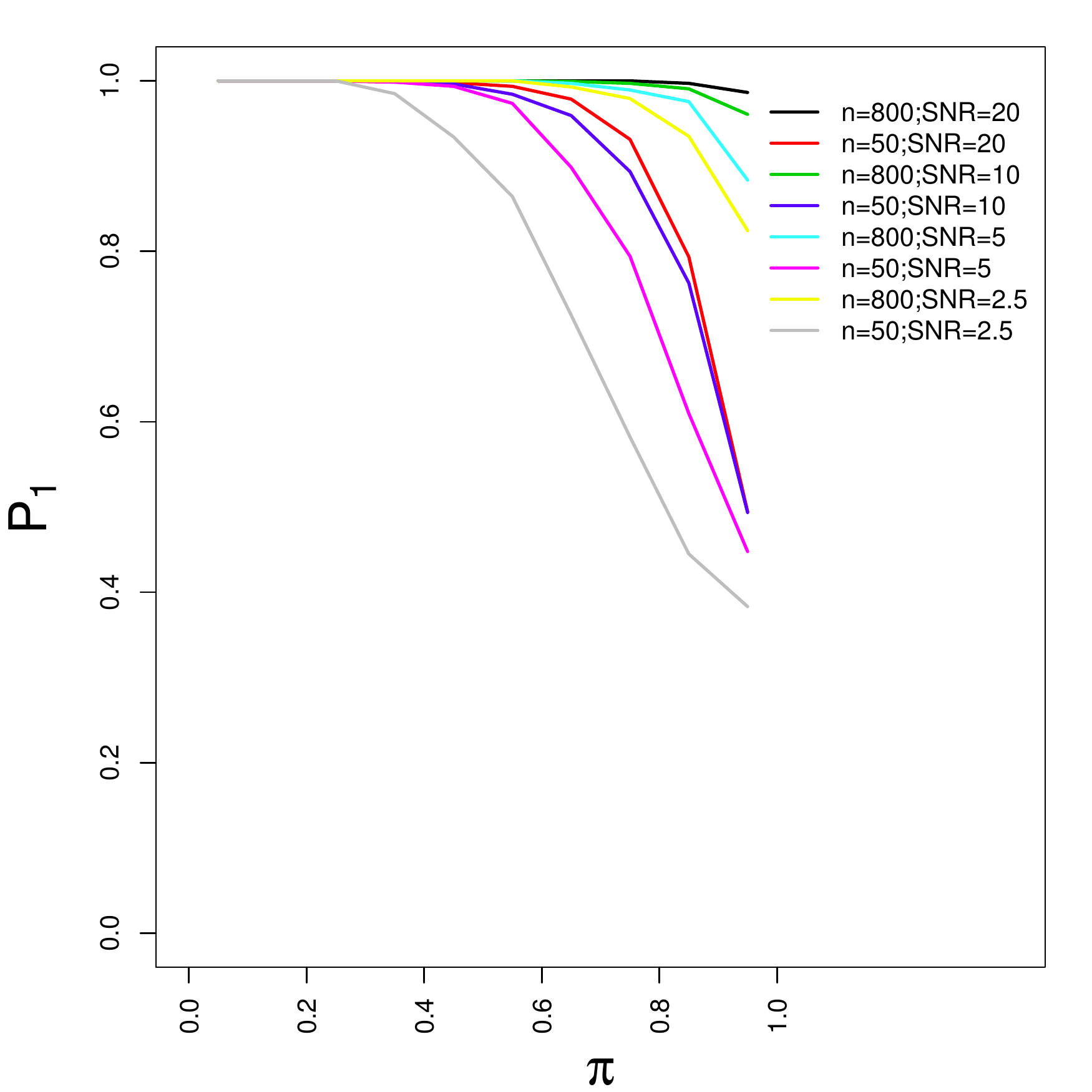}}
  	\subfigure{\includegraphics[scale=0.4]{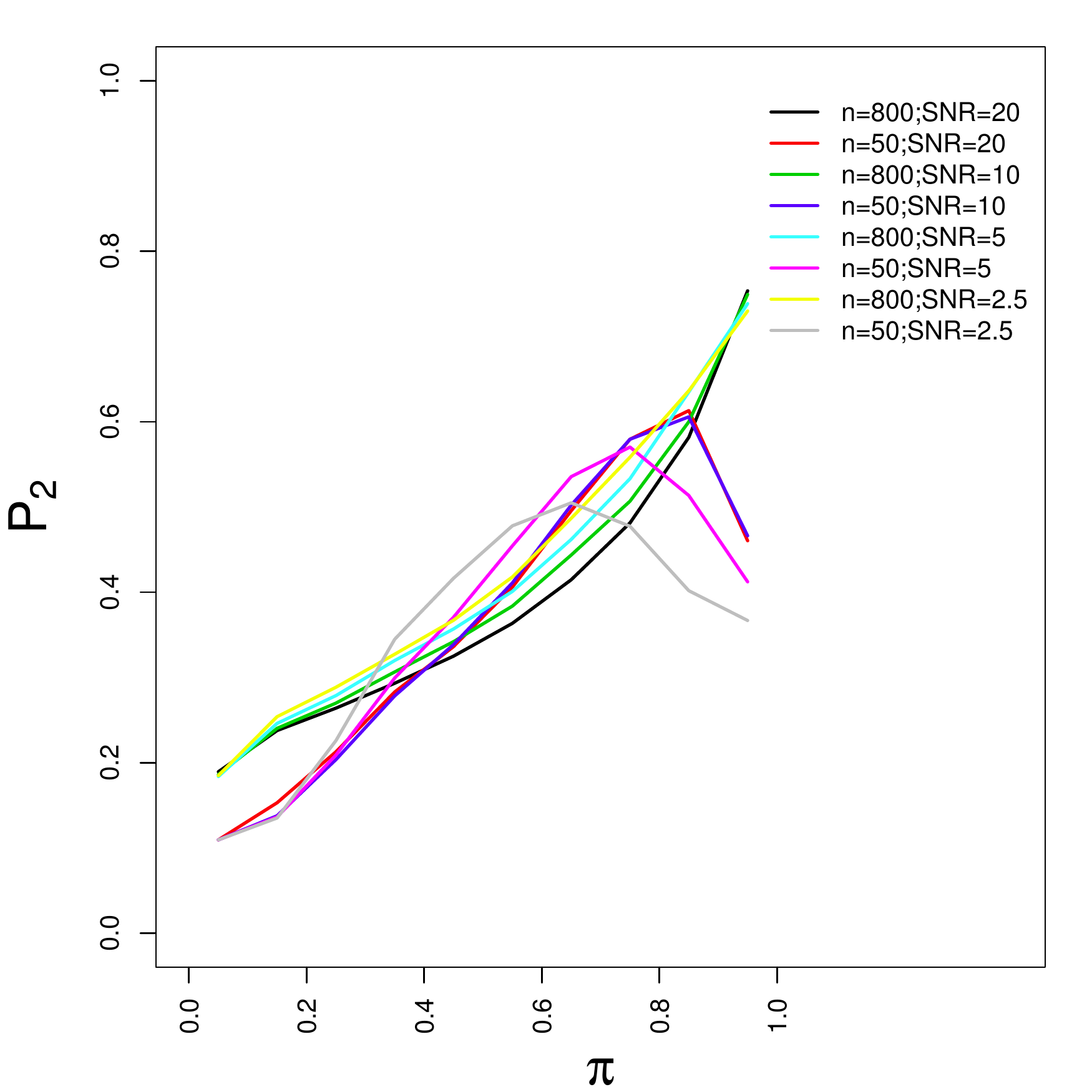}}

   \subfigure{\includegraphics[scale=0.4]{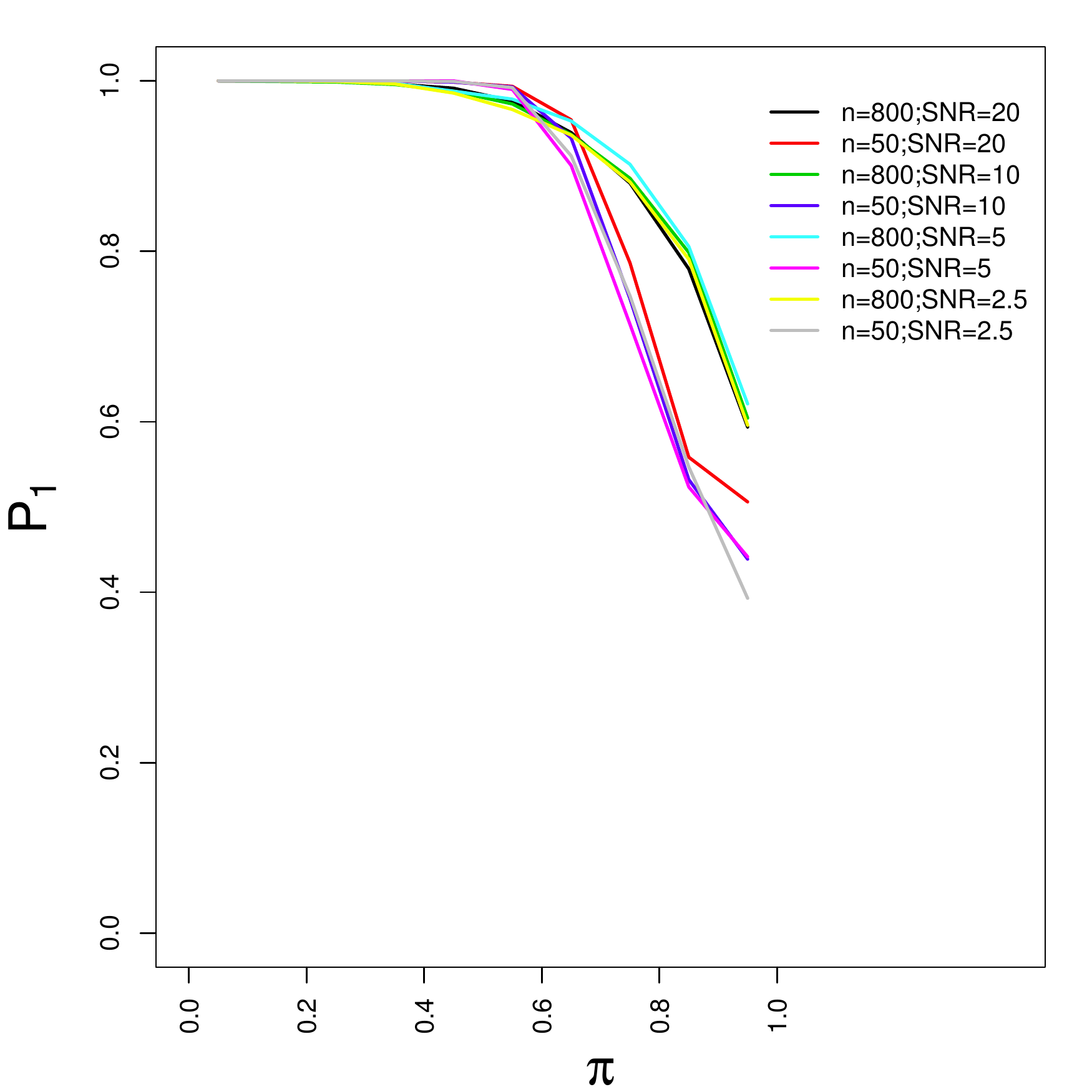}}
	   \subfigure{\includegraphics[scale=0.4]{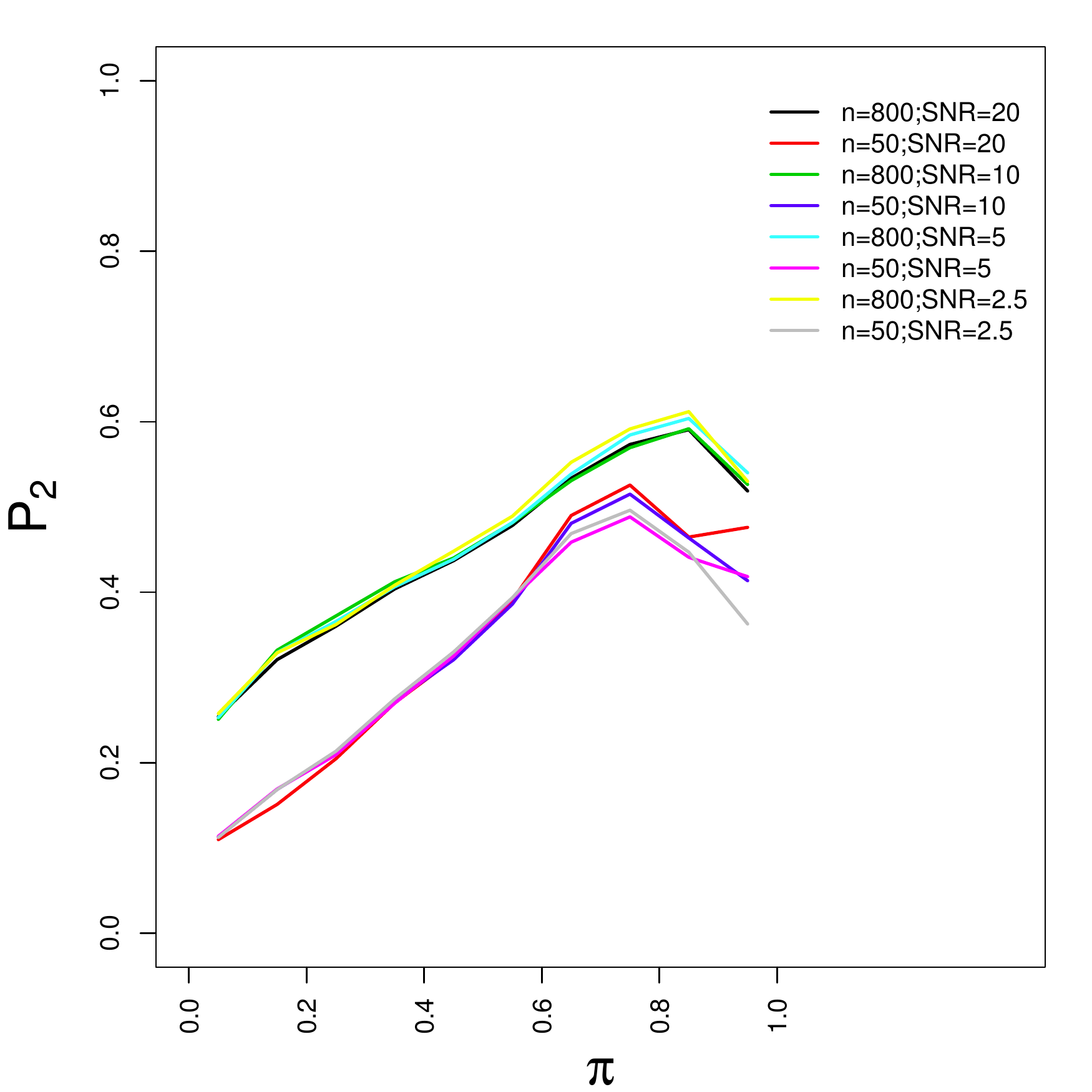}}
	  \caption{Mean of $P_1$ and $P_2$ calculated with different values of $\pi$. First row shows results of 1D/case 1, and second row presents results of 1D/case 2.}
\label{fig:1D sim path fig}
\end{figure}

Each simulation setting is repeated 100 times, and the average and the standard deviation of RMSE, $P_1$ and $P_2$ are reported in Table \ref{tab:1D Tab1} (for case 1 of $X$) and in Table \ref{tab:1D Tab2} (for case 2 of $X$). In each simulation scenario, the minimum of RMSE and the maximum of $P_1$ and $P_2$ are highlighted in bold. In all simulation scenarios, FuDoS outperforms the competing methods in prediction for a range of $\pi$. The best range of $\pi$, in the sense that it produces comparably small values of RMSE tends to fall as the sample size $n$ and $SNR$ becomes smaller. Specifically, when $n=800>>p=128$, the range of $0.35\leq \pi \leq 0.85$ results in very similar prediction results, and the smallest RMSE is achieved when $0.65\leq \pi \leq 0.85$. While when $n=50<<p$, $\pi>0.75$ generates poor prediction results, and the smallest RMSE is produced where $0.55\leq \pi \leq 0.65$.

We now discuss the selection results. Although WLasso is designed for producing sparse results, it does not exactly allow $\beta(t)=0$, so we did not measure $P_1$ and $P_2$ for WLasso. FLiRTI seems to perform well in selection as the produced $P_1$ (true positive rate) is very close to 1 in all simulation cases. However, the selected subset from the FLiRTI method tends to include many false segments as indicated by its fairly small values of $P_2$. The FuDoS method tends to identify the true subset more often than FLiRTI method as seen from its higher values of $P_2$. For instance, in case 2 of $X$, with $n=800$ and $SNR=20$, FuDoS produces $P_1=.99$ and $P_2=.44$, with $\pi=.45$, but FLiRTI yields $P_2=0.32$, with the similar value of $P_1=.99$. There is only one case (case 1 of $X$ with $n=800$ and $SNR=10$) that FuDoS emcompasses more false segments. To effectively visualize the selection performance of FuDoS, we plot the average of $P_1$ and $P_2$ for different values of $\pi$ in Figure \ref{fig:1D sim path fig}. It is not surprising that $P_1$ approaches to 1 as $\pi$ approaches to 0. Selection performance is related to both $n$ and $SNR$, but the effect of $n$ seems to be stronger.

\subsection{Analysis of gasoline data}
\label{sec:1D applications}
\begin{figure}
	\begin{center}
	  \subfigure[]{\includegraphics[scale=0.25]{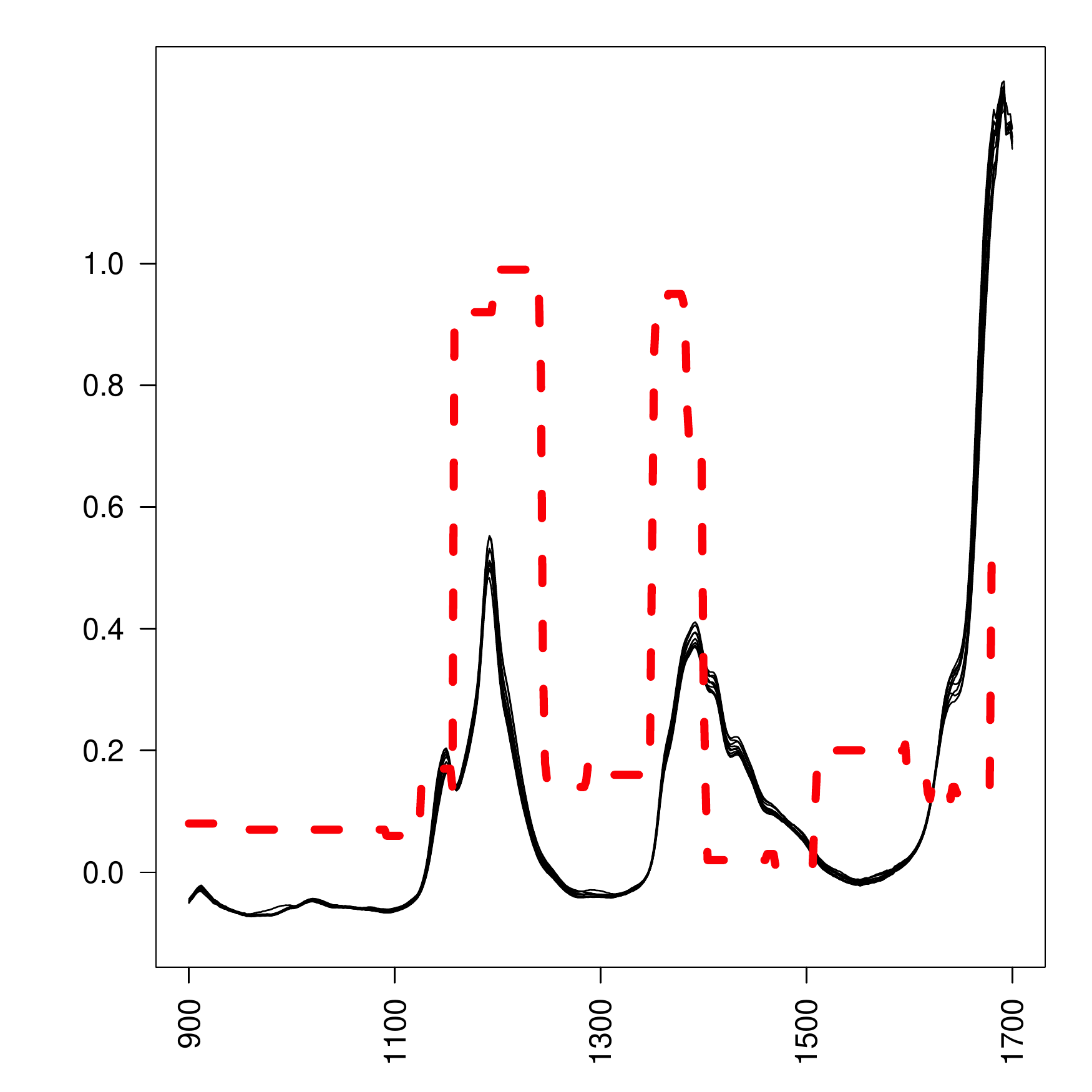}
	  \label{fig:gasolineFig1}}
	  	\subfigure[]{\includegraphics[scale=0.25]{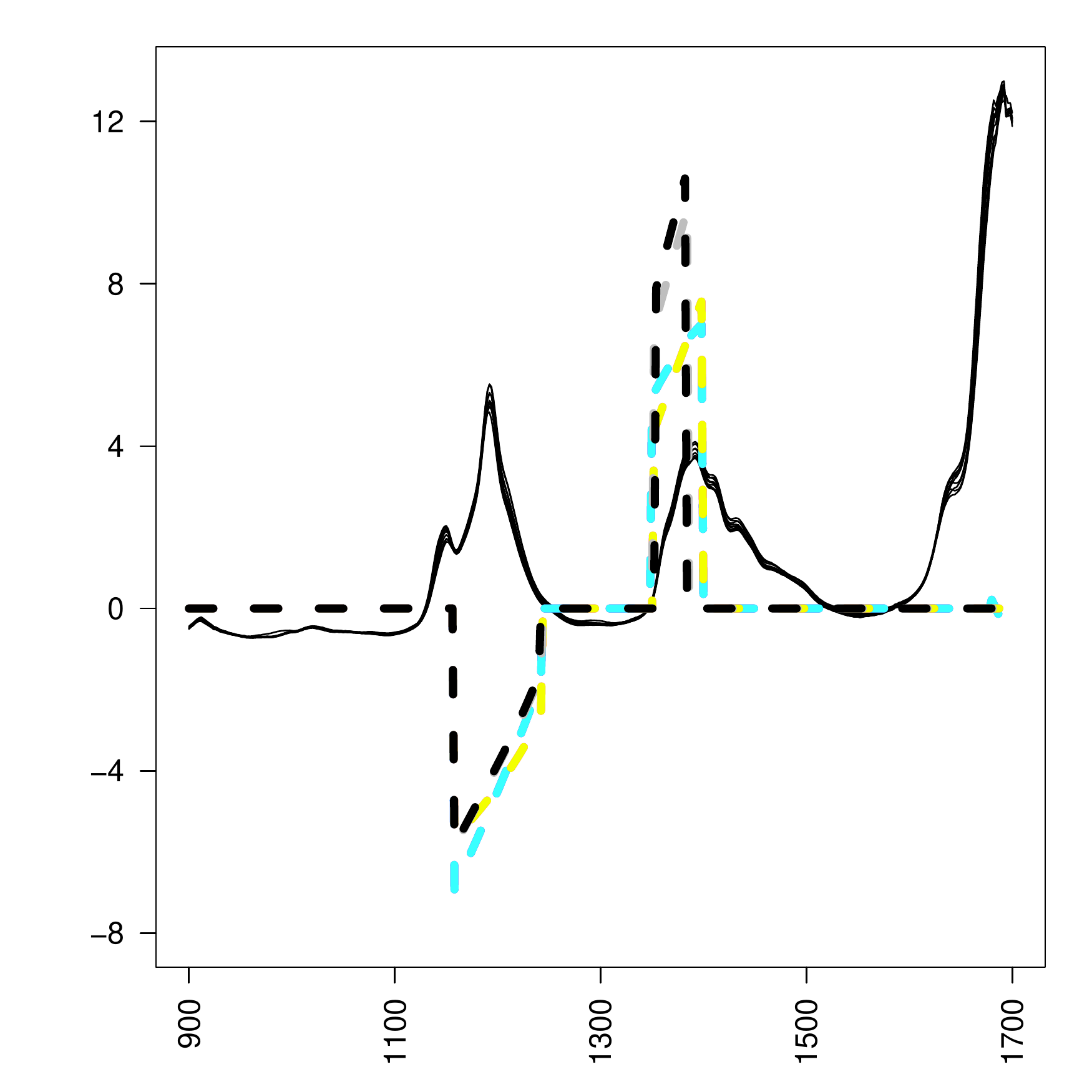}
	  \label{fig:gasolineFig2}}
   \subfigure[]{\includegraphics[scale=0.25]{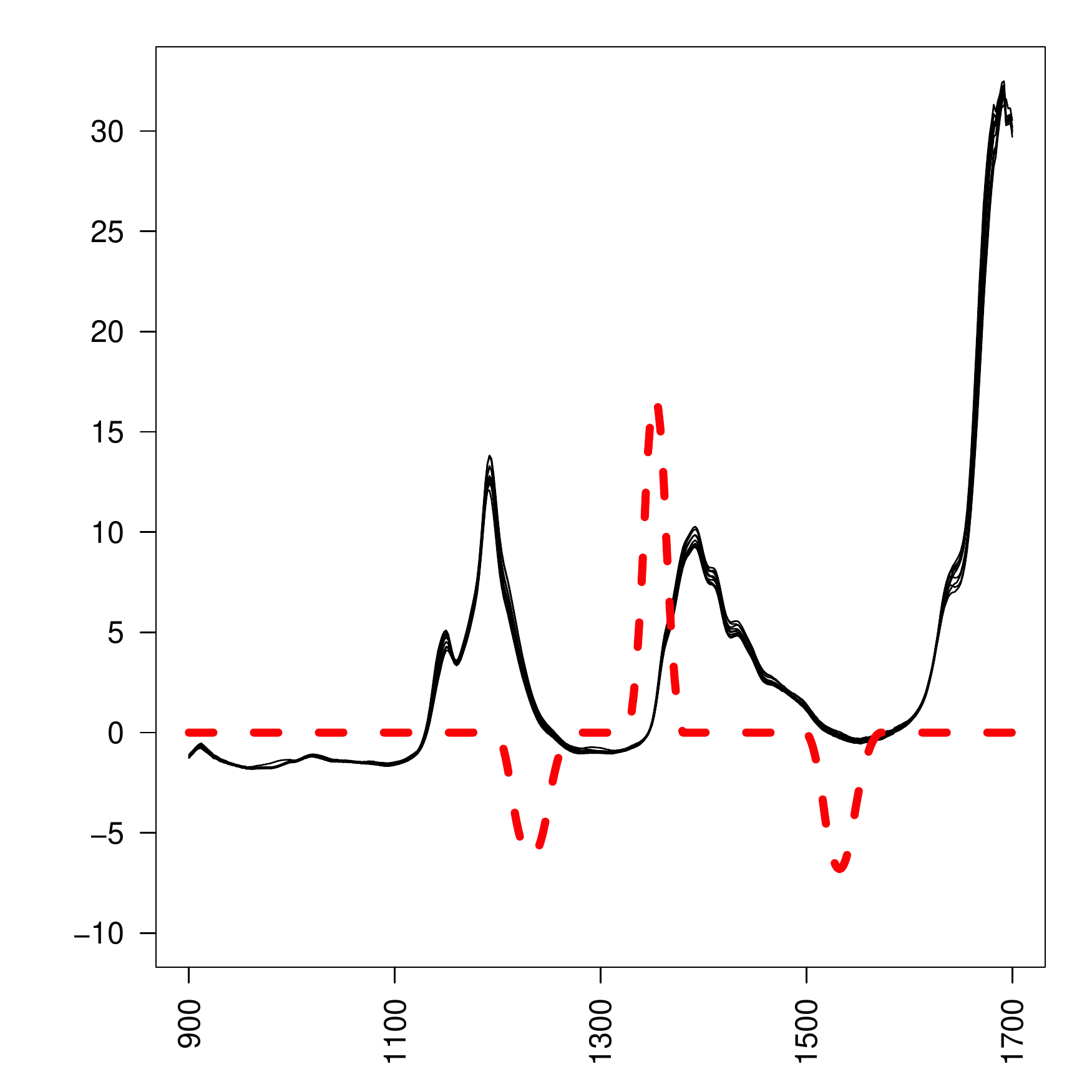}
	  \label{fig:gasolineFig3}}
	 \end{center}
			\caption{Analysis of gasoline data: (a) estimated maximum of selection probability (in red dashed). (b) estimated regression coefficient obtained by FuDoS for different values of $\pi=.35,.45,.55,.65,.75,.85$ (in different colour dashed). (c) estimated regression coefficient obtained by FLiRTI (in red dashed).}
\label{fig:gasoline fig}
\end{figure}

The gasoline dataset used in this article is obtained from the R-package \textit{refund} \citep{Rpackage:refund}. It contains the spectra of 60 gasoline samples, and for each of these samples, one observes the octane number. Each spectrum is measured by diffuse reflectance ranging from 900 nm to 1700 nm, and is digitized at 401 equi-spaced points. Black lines in Figure \ref{fig:gasolineFig1} illustrate 10 randomly selected spectra. Assuming that only some subregion of the spectra have a relation with the octane number, the goal is to identify such region of $\beta(t)\neq 0$ using the proposed method. As we did in the simulation study, we randomly subsample half of the dataset 100 times, returning 100 sets of $X^{{\rho,c}}_{\mathcal{J}^*}$, for each ${\rho,c}\in B$, where we use $B=(A\times C)$, $A=\left\{0.01,0.02,0.03,0.06\right\}$ and $C=\left\{0.01\right\}$. Through subsampling, we estimate maximum of selection probability and display it in Figure \ref{fig:gasolineFig1} in a dashed red line. The regression coefficient $\beta_{[l]}$ is estimated using penalized B-splines methods, with the smoothing parameter selected by GCV, and the smoothing parameters are assumed to be the same for all $l=\kappa_1,...,\kappa_K$. Figure \ref{fig:gasolineFig1} reveals that FuDoS identifies two parts of the spectra, roughly 1150-1250 nm and 1320-1370 nm, being related to the octane number with high probability. We now estimate $\beta(t)$ using the estimated stable subdomain $\hat{\mathcal{X}}^{\pi}_{stable}$, for different values of $\pi=.35,.45,.55,.65,.75,.85$, and display them in Figure \ref{fig:gasolineFig2}. For comparison, we fit the FLiRTI method to the dataset, as given in Figure \ref{fig:gasolineFig3} in a dashed red line, where the involved two tuning parameters were chosen via a 5-fold CV, and the used two derivative orders were $d=0,3$. It is shown that the octane number is negatively related to the spectra between 1200-1270 nm, but has a positive association near 1350 nm, which is consistent with the result from FuDoS. There is one part in \ref{fig:gasolineFig3} not identified by FuDoS, i.e., wavelengths between 1500-1550 nm are selected by FuDoS with probability less than 0.2.

\begin{figure}
	\begin{center}
	  \subfigure[FuDoS]{\includegraphics[scale=0.25]{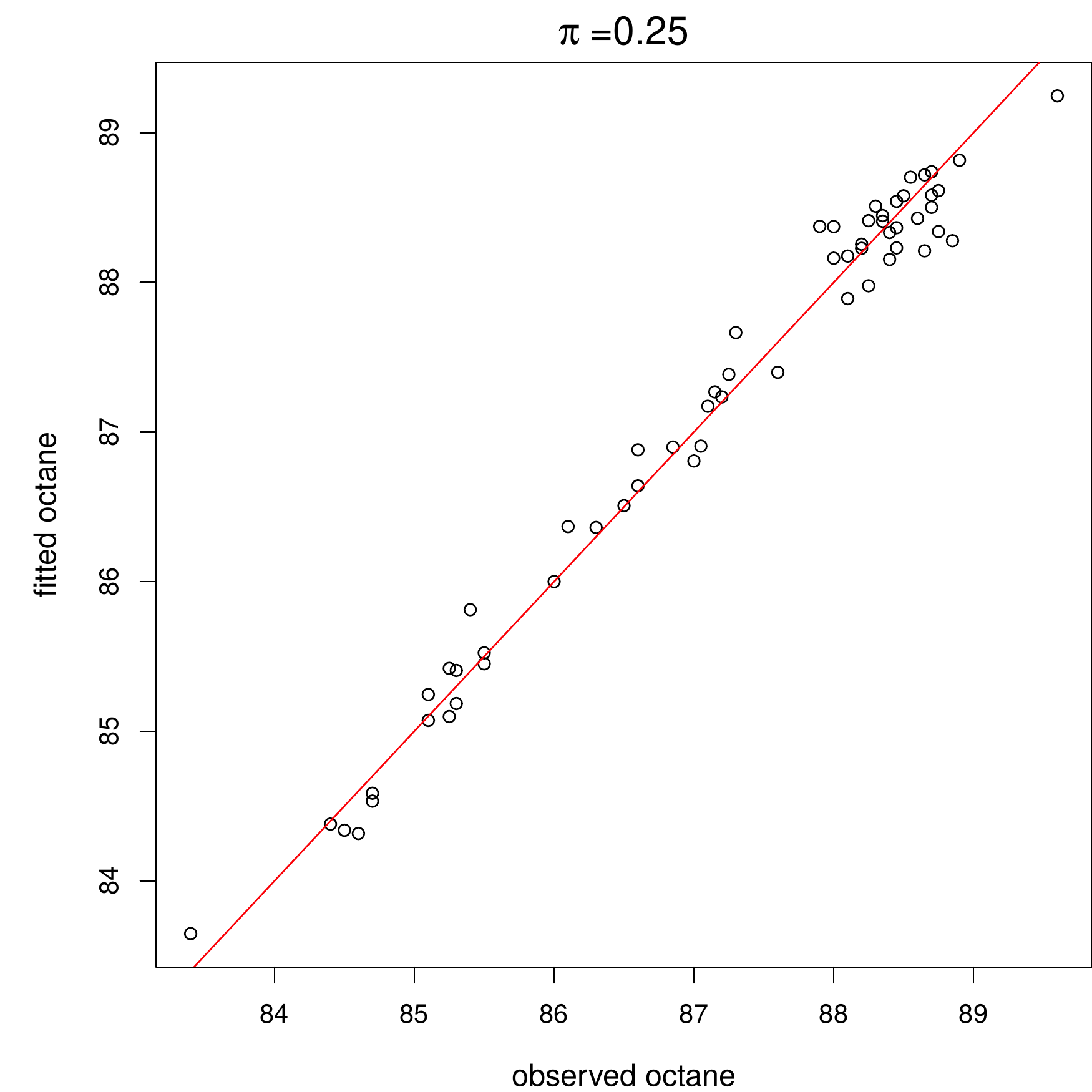}
	  \label{fig:gasolineFig4}}
	  	\subfigure[FuDoS]{\includegraphics[scale=0.25]{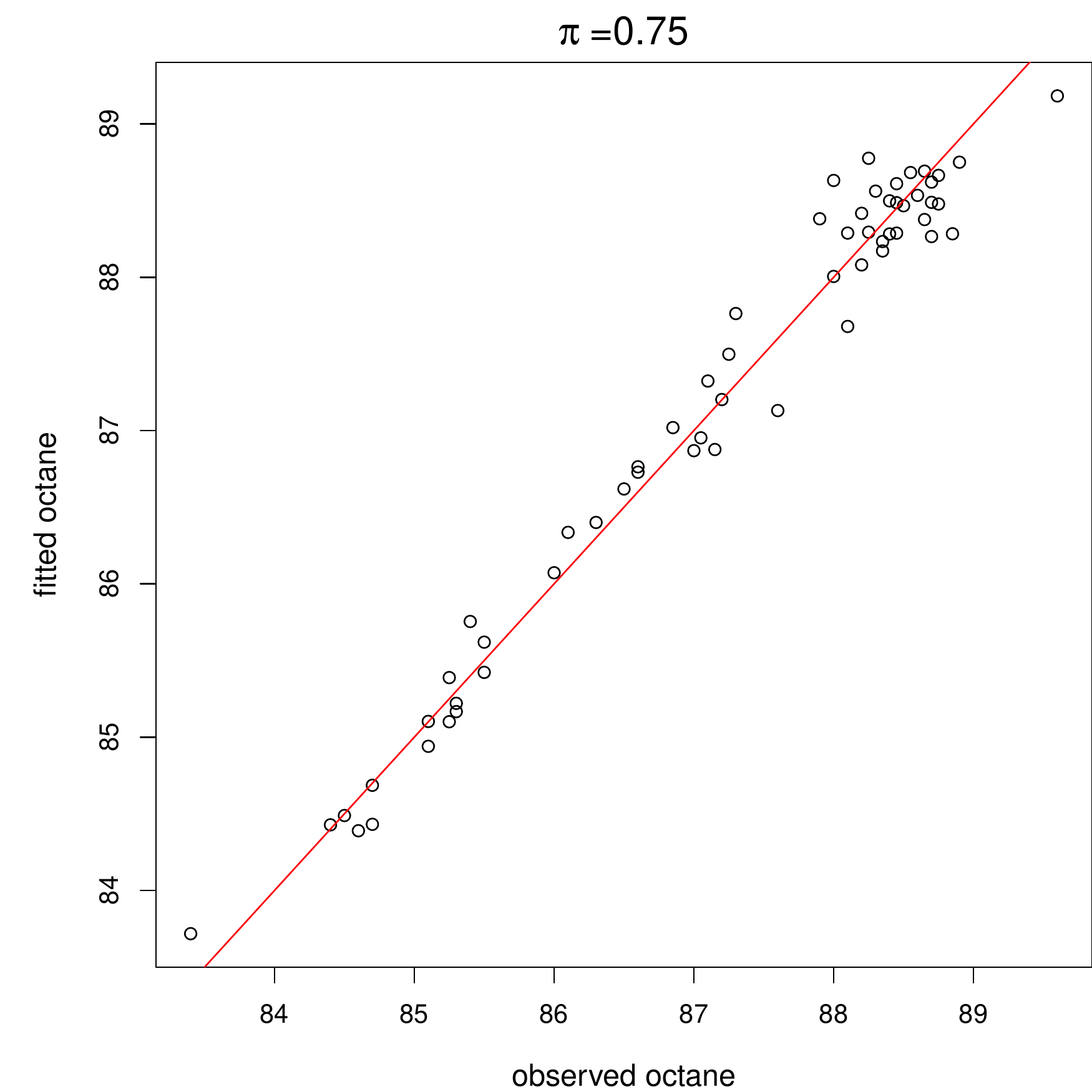}
	  \label{fig:gasolineFig5}}
   \subfigure[FLiRTI]{\includegraphics[scale=0.25]{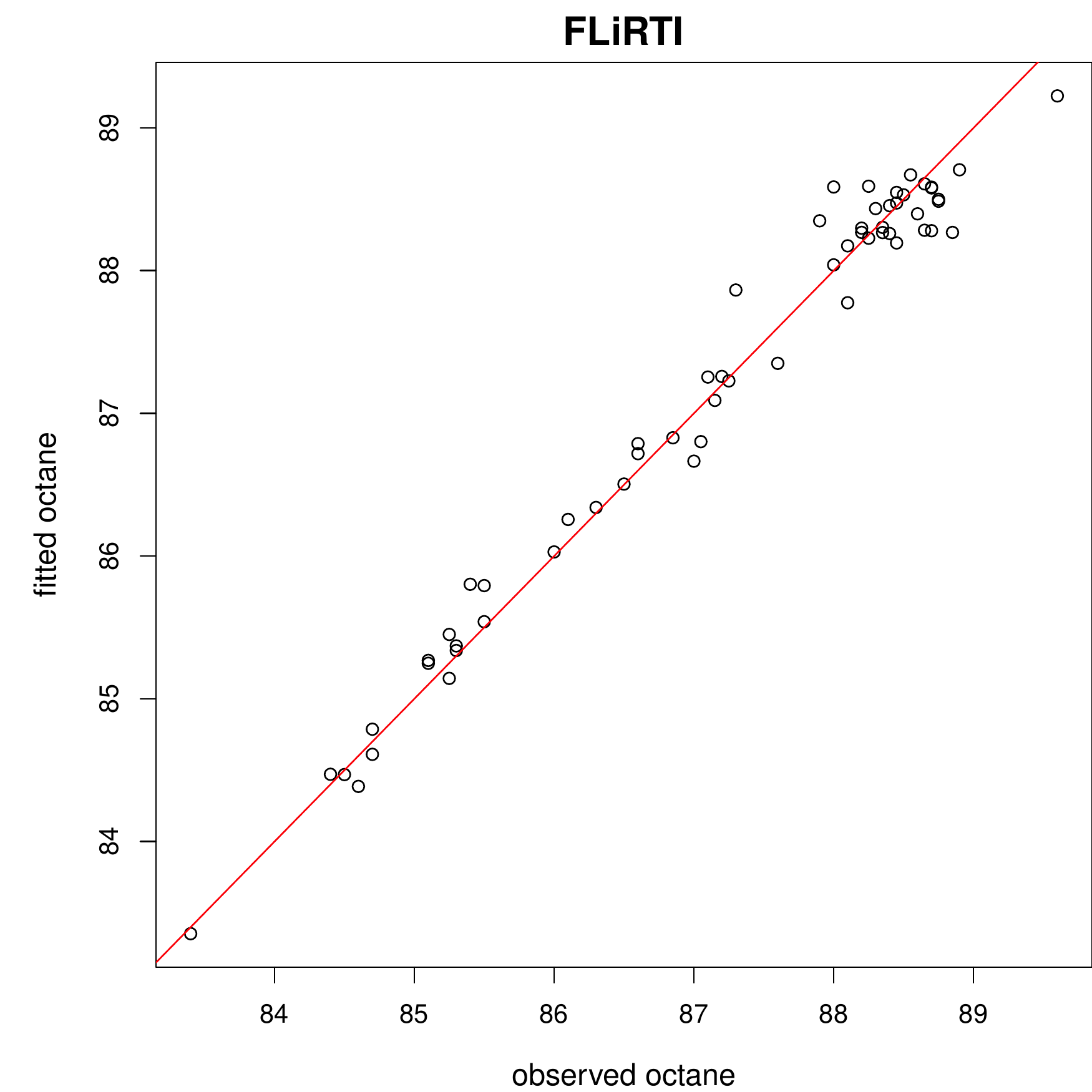}
	  \label{fig:gasolineFig6}}
	 \end{center}
			\caption{Observed (x-axis) versus predicted (through 10-fold cross-validation) octane number (y-axis).}
\label{fig:gasoline pred fig}
\end{figure}

Next, we exploit 10-fold cross validation to test the predictive ability of the two methods. Specifically, each sample fold of 10\% observations is left out as a validation set, the rest is used to train the model, and the prediction is performed on the observations that have been left out. Repeating this procedure for each sample fold, we aggregate the predicted values, and compute the predictive R-square, defined as $1-\sum(Y_i-\hat{Y}_{-j,i})^2/\sum(Y_i-\bar{Y})^2$, where $\hat{Y}_{-j,i}$ is the predictive value of $Y_i$, based on the model fitted from training samples except samples in $j$th sample fold. The predictive R-square produced by FuDoS is 0.97-0.98 for a range of $0.1<\pi<0.9$, and the R-square yielded from FLiRTI is 0.98. The plots of the original versus predicted octane number are shown in Figure \ref{fig:gasolineFig4}.

\end{document}